# Timing uncertainty in collective risk dilemmas encourages group reciprocation and polarization

**Elias Fernández Domingos[1,2,3,*], Jelena Grujić[1,2,*], Juan C. Burguillo[3], Georg Kirchsteiger[4], Francisco C. Santos[5,6,2,‡], and Tom Lenaerts[1,2,‡]**

[1]AI lab, Computer Science Department, Vrije Universiteit Brussel, Pleinlaan 9, 3rd floor, 1050 Brussels, Belgium

[2]MLG, Département d'Informatique, Université Libre de Bruxelles, Boulevard du Triomphe, CP 212, 1050 Brussels, Belgium

[3]Department of Telematic Engineering, University of Vigo, 36310 Vigo, Spain

[4]ECARES, Université Libre de Bruxelles, Av. Roosevelt 42, CP 114, 1050 Brussels, Belgium

[5]INESC-ID and Instituto Superior Técnico, Universidade de Lisboa, IST-Taguspark, 2744-016 Porto Salvo, Portugal

[6]ATP-group, 2744-016 Porto Salvo, Portugal.

* equal contributors

‡ corresponding authors

# Abstract

**Human social dilemmas are often shaped by actions involving uncertain goals and returns that may only be achieved in the future. Climate action, voluntary vaccination and other prospective choices stand as paramount examples of this setting [1–4]. In this context, as well as in many other social dilemmas, uncertainty may produce non-trivial effects [5]. Whereas uncertainty about collective targets and their impact were shown to negatively affect group coordination and success [6–8], no information is available about *timing uncertainty*, i.e. how uncertainty about when the target needs to be reached affects the outcome as well as the decision-making. Here we show experimentally — through a collective dilemma wherein groups of participants need to avoid a tipping point under the risk of collective loss[8] — that timing uncertainty prompts not only early generosity[9] but also polarized contributions, in which participants' total contributions are distributed more unfairly than when there is no uncertainty. Analyzing participant behavior reveals, under uncertainty, an increase in reciprocal strategies wherein contributions are conditional on the previous donations of the other participants, a group analogue of the well-known Tit-for-Tat[10,11] strategy. Although large timing uncertainty appears to reduce collective success, groups that successfully collect the required amount show strong reciprocal coordination. This conclusion is supported by a game theoretic model examining the dominance of behaviors in case of timing uncertainty. In general, timing uncertainty casts a *shadow on the future*[11] that leads participants to respond early, encouraging reciprocal behaviors, and unequal contributions.**

## Introduction

Public good games (PGG) provide abstractions of many real-world problems wherein personal and short-term interests of multiple players are in conflict with societal, long-term interests [12]. Participants in such games can contribute voluntarily to a common good, which can, once established, be accessed without restrictions by all, thus even those that did not contribute. Rational selfish behavior stipulates that it is best to not contribute; yet such decision would be detrimental as a group is better off when all contribute. These games not only serve as a good model for many for social public benefits (e.g., social security, retirement funds), but are also recurrent in many other collective endeavors, from group hunting [13,14] to public health [3,15–18] and sociopolitical processes like climate change [1,6,19–22].

We place our experimental within the collective risk dilemma (CRD)[8], a variant of a PGG, where a threshold makes the outcome of the game non-linear, and the collective benefit uncertain and only achievable in the future[23–25]. This model has been adopted to address the complexity pertaining decision-making under climate dilemmas [8,9,21,26], but its significance is general enough to be of interest to a broad range of Human endeavors, such as costly signaling, voting or petitioning. At the start of the game, participants are each given an endowment ($E$), and they must decide whether to contribute, up to a predefined amount, to the common good over a fixed number of rounds. If the joint contributions of all the participants over those rounds are equal or above a certain threshold, then the disaster is averted, and they receive as a reward the remainder of the endowment (hence the dilemma). On the contrary, if the target is not reached, there is a probability that a disaster may occur, resulting in economical loss for all the participants (they lose the remainder of their endowment). This is modelled by a risk parameter and both experiments and theoretical analysis show that people only tend to contribute to avoid the disaster if they perceive the risk to be high[8,20–22,26–29]. Moreover, even when the risk is high, theoretical models indicate that players tend to delay their contributions when the moment of disaster is known[21,30].

Both threshold PGG and CRD make strong assumptions about what is known in the game: Each participant knows from the start how much they need to acquire collectively to reach the target and how much time they have to achieve this. Yet in real-world scenario's the amount as well as the timing when it has to be achieved may not be certain, as they are based on predictions and thus inherently suffer from uncertainties. Prior work on uncertainty about what amount (threshold) should be achieved in PGG [5,31] and CRD [6,7] has shown that the level of cooperation, i.e. the willingness to contribute in both games, is negatively affected, yet no insights exist on how timing uncertainty affects the decision-making process.

To answer this question, three experimental treatments are performed here. First, as a control treatment (NU – no uncertainty), we investigated the behavior of groups of 6 participants, wherein each can contribute 0, 2 or 4 EMUs (Experimental Monetary Units) at each of the 10 rounds of the experiment ($m_0 = 10$). When the group does not reach the target contribution of 120 EMUs by the $10^{th}$ round, they risk losing the remainder of their initial endowment (40 EMUs) with a 90% probability. NU, thus, repeats the work of Milinski et al. [8], but *without* the climate change framing, which avoids possible cultural effects due to climate awareness, while enabling the generalization of our conclusions to other problems captured by the CRD. Conversely, in the second treatment (LU – low uncertainty), participants did not know exactly when the experiment would finish. They were told the experiment lasts on average 10 rounds and that from round $m_0 = 8$ there was a possibility that the game ends (see Supporting Information (SI) for instructions): a 6 faces dice was thrown, and the game would continue if the result was higher than 2 (i.e., the game ends with a probability of $w = 1/[(10 - m_0) + 1] = 1/3$). If the game continues, the same process is repeated at the end of each round, until the dice result indicated the end of the experiment. The game can thus stop early at round 8 but also continue multiple rounds after round 10. Finally, for the third treatment (HU – high uncertainty) we increased the uncertainty, i.e., the variance of the distribution of final rounds, by making $m_0 = 6$ (and $w = 1/5$ – we throw a 10 faces dice in this case). Importantly, we made sure that all participants are clearly informed, in every treatment, that the average number of rounds is 10 (their understanding was tested before starting the experiment). Note that in all three treatments, fair behavior would be to contribute in total half of one's endowment (*F=E/2*), as it would ensure that everyone has the same, maximized gains and that the target is met.

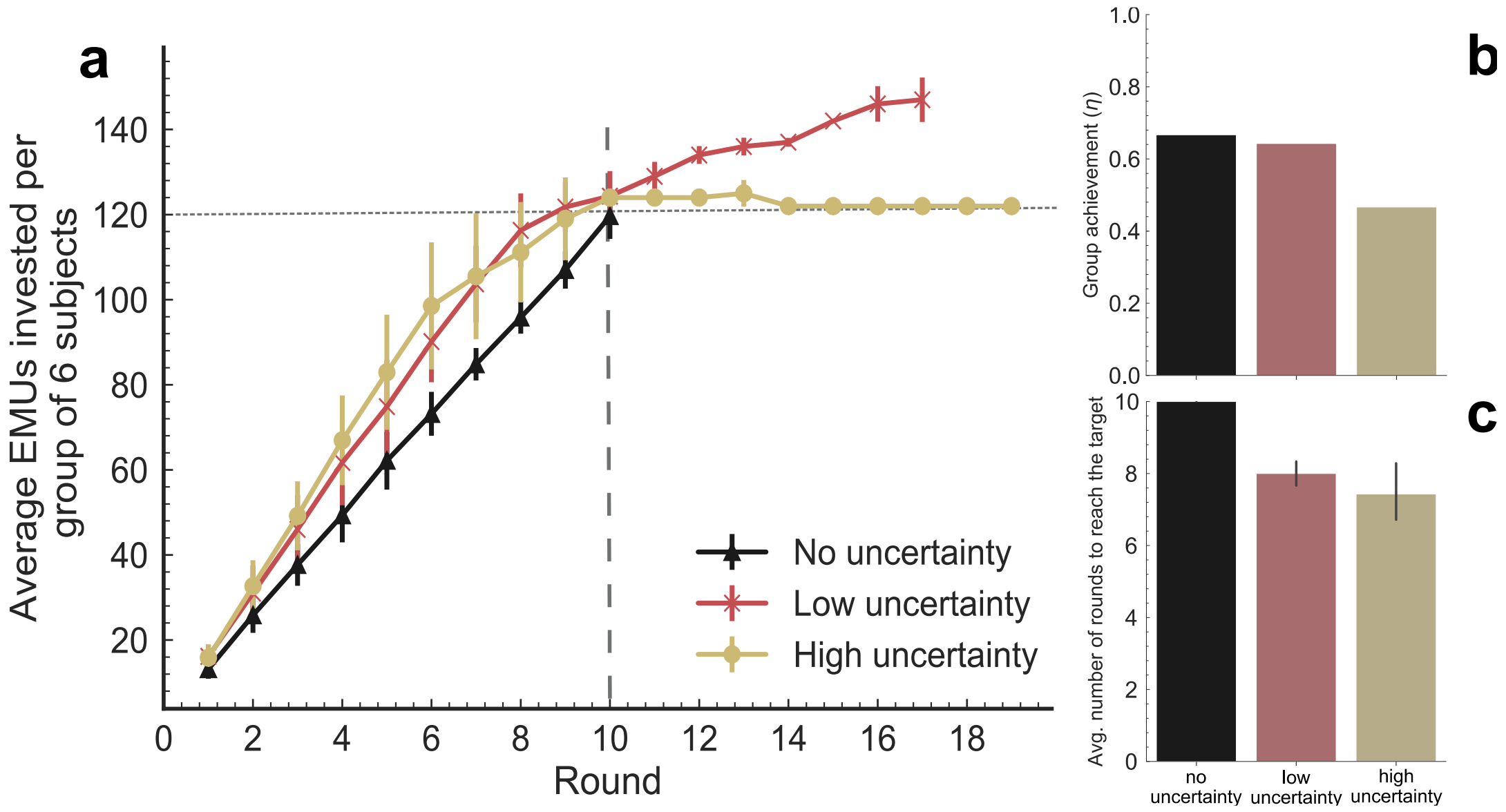

**Figure 1. Average investment in the public account (a), average group achievement (b) and average number of rounds used to reach the target (c).** Panel (a) shows the average investment per round and per group of 6 players for each treatment**.** For NU (black lines and triangles), the groups invested per round on average the minimum amount required to reach the target by the end of the game (12 EMUs per round). Treatments LU (red lines and stars) and HU (yellow lines and circles) assume low and high timing uncertainty, respectively, even though the game would on average end at 10 rounds. Our results suggest that the higher the uncertainty on the timing of the final round, higher the amount contributed at early stages of the game. As a result, when groups succeed in reaching the collective threshold, they do so in a smaller number of rounds (see Panel (c)). However, these early investments did not lead, on average, to higher collective success. Instead, in panel (b), we can see that the group achievement of HU is lower than NU and LU.

## Results

In the absence of any timing uncertainty (NU, black lines and bars in Figure 1), the experimental results show that groups can reach the target amount, and that their investments per round follow closely the minimum required to reach the target by the end of the 10th round (Figure 1a). Figure 1b shows that more than 65% of the groups were successful (average group achievement, $\eta$). These groups nonetheless reached the targeted amount only by the last round (Figure 1c), and those that did not, were always very close to it, indicating a failure in the final coordination. On the contrary, whenever timing uncertainty is present (LU and HU), individuals tend to contribute earlier in the game, with the amount contributed per round increasing with the uncertainty on the game timeframe (see Figure 1a). Figure S1 in SI, provides similar conclusions when examining the average donations per round and per group.

The motivation behind each individual's decision to contribute early is certainly diverse. In LU, the investments are very close to the required donation per round for a game with only 8 rounds. Indeed, as shown in Figure 1c, groups that met the target in LU, did this by round 8 on average, while the fraction of groups that met the target did not change significantly (Figure 1b, see Methods for statistical test). This indicates that participants attempted to reduce the uncertainty of the game (and the risk associated) by investing earlier. Yet, in HU, the average investments were lower than those required to reach the target in 6 rounds (see Figure 1a). In this third treatment, groups only managed to reach the target on average by round 8 (see Figure 1c), which indicates that uncertainty affects behavior (see also Figure S2 in SI) in a way that is subtler than simply understanding LU and HU treatments *as if* players would be playing a distinct game with less rounds. In other words, under timing uncertainty, individuals behave, on average, in a risk-averse fashion, reacting earlier to the collective dilemma, and finding it unlikely that the game would end after the average number of rounds (10, in this case).

Independent of when they try to reach the target, if every participant invests half of his or her $E$ then the target is reached, and they all go home with the same gains. This *fair share* $F$ can be achieved by accumulating different combinations of 0, 2 or 4 EMUs over the different rounds. Indeed, we observe in the investments per round for the NU experiment that most participants giving $F$ do not do this by giving 2 in every round. Only 15% (11 out of 72) of all participants giving $F$ and 20% (10 out of 48) of the successful ones do this by always giving what could be the locally fair share. Fairness in this game is thus defined at the game-level and not the round-level.

Now, when considering, within each group, the cumulated investments over all rounds ($C$), one can observe that individuals react to uncertainty by either giving more or less than from $F$ (see Figure 2a),. This suggests that the presence of timing uncertainty not only leads to earlier investments, as mentioned earlier, but also promotes the emergence of polarized reactions and unequal contributions among participants, with more players' C deviating from $F$ than when there is no uncertainty. This observation is confirmed in HU (see Figure 2a), where the prevalence of unfair contributions increases further when compared with NU and LU. This polarized pattern may suggest a co-existence of risk-averse and risk-seeking individuals [32,33], depending on whether individuals base their choices on a number of rounds below or above the average of 10 rounds.

The same cumulative data also reveals the shift towards early investments when considering players' donations in the beginning and end of the game (see Figure 2b). Our results show how participants distribute their donations over time, by dividing the game into two halves (from $[1, m_0/2]$ and $(m_0/2, end]$, i.e., half means 5, 4 and 3 for NU, LU and HU respectively). The fraction of players whose C is more, equal or less than half of the *fair donation*, i.e., $F/2$, is shown for the groups that reached the target. In the treatments with uncertainty (LU and HU), the fraction of $C > F/2$ players in the 1st half is significantly higher than that of the 2nd half, which means that participants invest earlier to reach the target. In contrast, in NU there is a slight increase in $C > F/2$ during the second half of the game. This may be related to players trying to compensate for *procrastination*, resulting in higher investments at the end. Moreover, when comparing the donations between players that met and did not meet the target (see Figure S3 in SI), the difference between the fraction of players that invest $C > F/2$ in the 1st half of the game grows with uncertainty. This highlights the importance of investing earlier and not *procrastinating* under the presence of timing uncertainty.

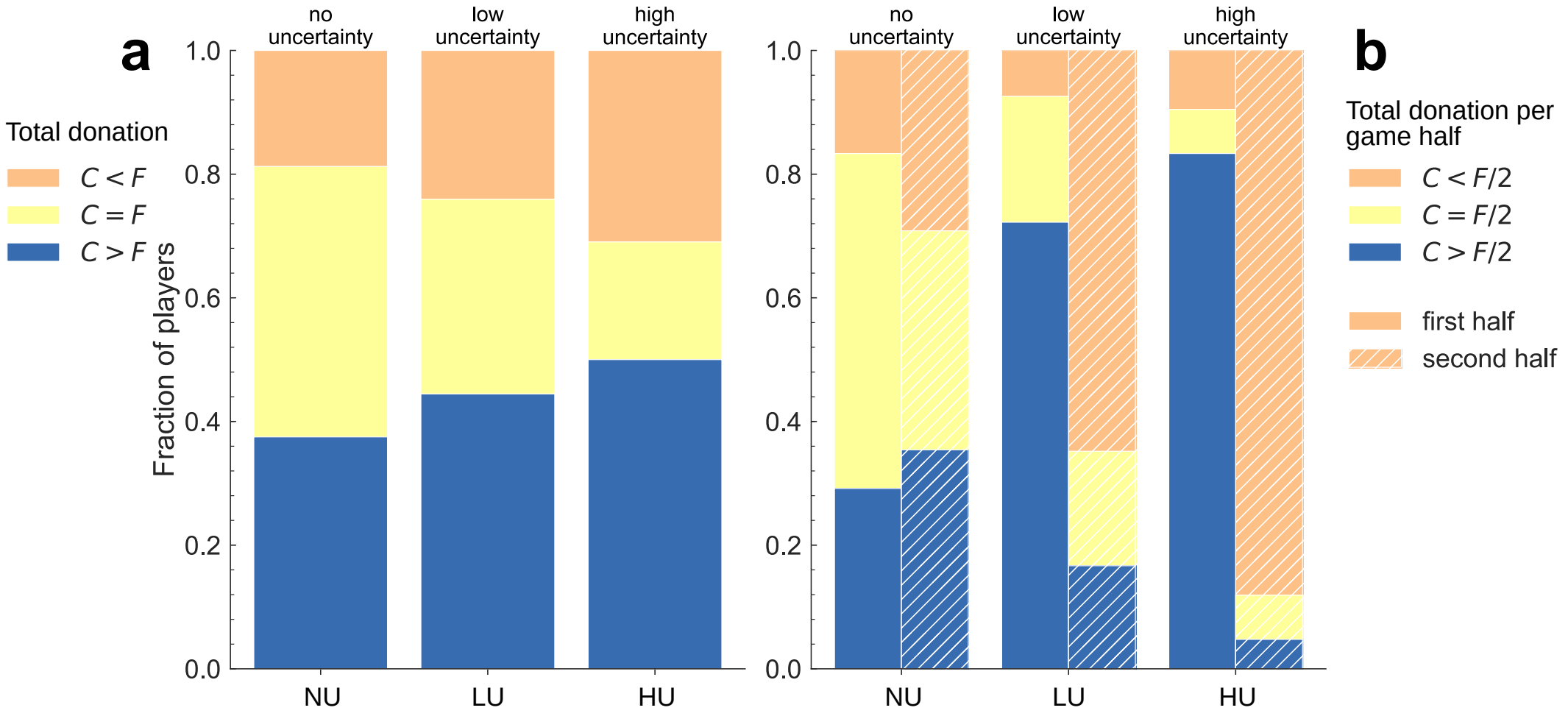

**Figure 2. Distribution of participant behavior and dependence on time.** Fraction of players within successful groups that invest less, equal or more than the *fair donation* $\boldsymbol{F}$ (if all participants of a group invest exactly $\boldsymbol{F}$, they reach the target) (a) and how these behaviors are distributed over the first and second halves (1 half means 5, 4, and 3 rounds for NU, LU and HU respectively) of the game (b). Panel a shows that as uncertainty increases, more players invest either more or less than $\boldsymbol{F}$, while the players that invest exactly $\boldsymbol{F}$ decrease. Panel b shows how this behavior is spread across the game, by showing the fraction of players that invest less, equal or more than $\boldsymbol{F/2}$ in each half of the game. The plot shows that the investments in the first half of the game increase with uncertainty, in comparison to NU.

Not only do participants tend to invest earlier in the presence of timing uncertainty, but their donations become dependent on what the group members did (see Figure 3), to the point that the predominant behavior in successful LU and HU resembles a group-level reciprocal behavior[34] or Tit-for-Tat (TfT)[10,35]. Such group reciprocal behavior, or *group conditional cooperation*, has also been observed in linear public good games without a threshold[36,37], and it has been identified experimentally as a beneficial strategy for climate action[38].

Although this behavior does not avoid free-riding (there is actually and increase as can be seen in Figure 2a), it promotes generosity among the participants and appears necessary for a successful outcome. In Figure 3, we can see that there is a positive correlation between the group donations and the average donation of the players in LU and HU (see correlations in Table S1 in SI) for the players that met the target. In contrast, the players that did not meet the target do not display the same conditional behavior or even to the same extent, indicating that reciprocity is used here by the participants to achieve "honest" coordination in the presence of timing uncertainty.

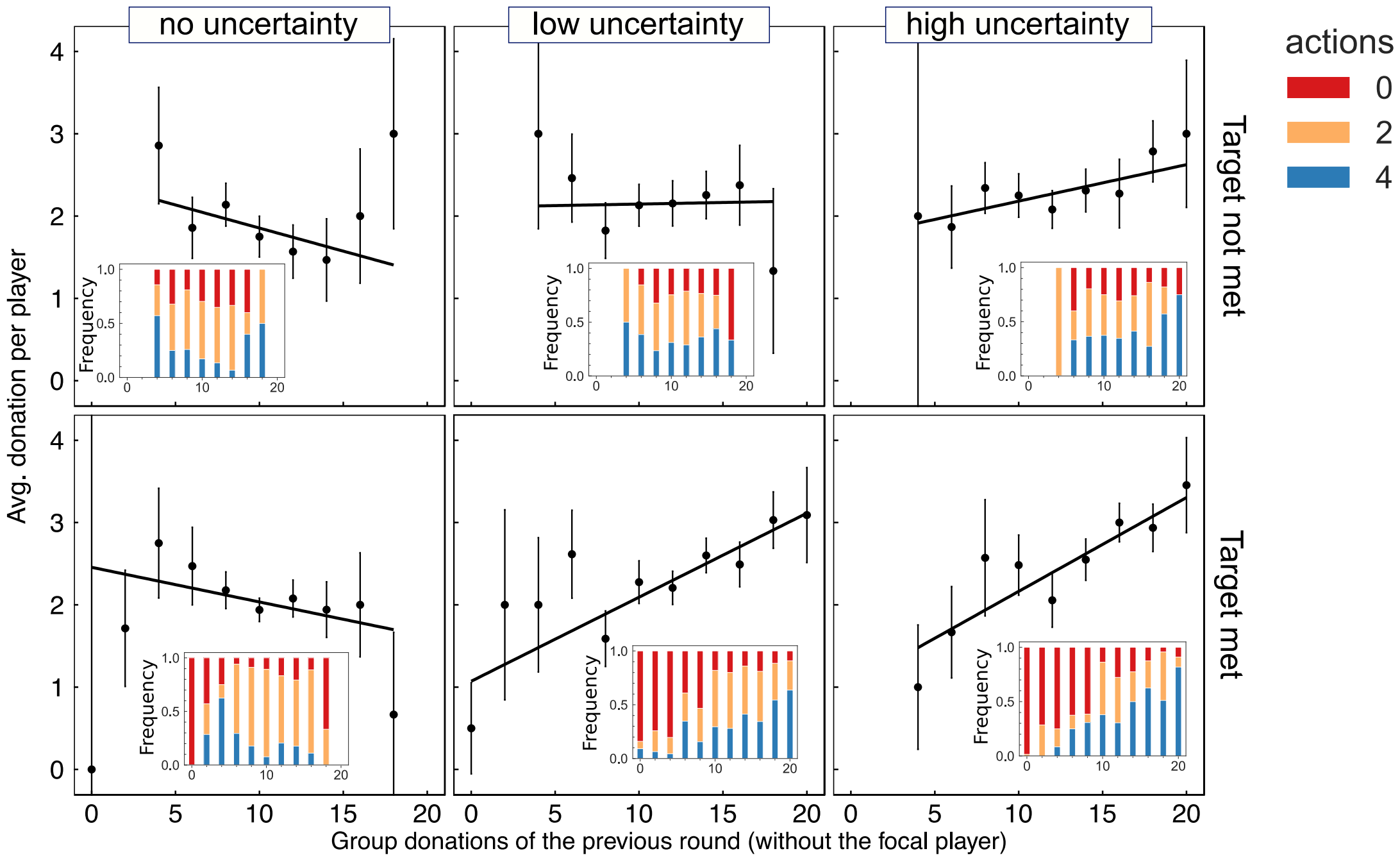

**Figure 3. Prevalence of group reciprocity for different uncertainty levels.** We show the average donation per player in function of the donations of the group members in the previous round (without the focal player). The plots are separated by treatment (columns) and by whether they met (True) or not (False) the target (rows). We fitted a weighted linear regression on each plot (see Methods). This analysis shows that there is no significant dependency on the group donations for NU, despite a slightly negative correlation factor for the players that reached the target, which indicates a compensatory behavior. However, there is a significant dependency on the group donations under uncertainty (LU and HU). Moreover, for LU, the groups that did not reach the target display a slight compensatory behavior, in contrast to the reciprocal behavior of those that did. Inside each plot, we show a subplot of the frequency of each action (0, 2 and 4) for the different group donations. These plots depict clearly how, for LU and HU, the frequency of action 4 increases with the group donation on the previous round, while action 0 increases when the previous donations were low. In comparison, action 2 is predominant when there is no uncertainty or when groups did not achieve the goal (LU-HU).

The non-trivial dynamics and behavioral ecology of switching from compensating to reciprocal behavior may be explained by a game theoretical model, that describes the behavioral dynamics through an evolutionary process (see Methods for a full description). Given the strategy profiles identified in Figure 3, human subjects mostly adhere to either unconditional investments or conditional strategies based on the level of investments in the previous round. For simplicity, we define here three unconditional heuristics or strategies, i.e. *always-2* (gives 2 in every round), *always-4* (invest gives 4 in each round), *always-0* (invest nothing throughout the game), and the two conditional ones, i.e. *compensator* (will invest 4 when the group members did not invest) and *reciprocal* (will invest 4 as long as the group members invest). All strategies will stop contributing once the collective target is achieved. We consider a population of individuals who may adopt one of these five strategies and revise their choices based on the relative success of each strategy [39] (see Methods for details). More advanced strategies could be considered, yet even with this baseline model, our experimental observations can be explained. Indeed, the model confirms that, under uncertainty, the reciprocal strategy prevails among those strategies that contribute to the collective good (see Figure 4a), while capturing also that the group achievement does not change significantly with uncertainty. Moreover, the model indicates that the *always-2* strategy is only stable when there is no timing uncertainty (see Figure S6 in SI). The presence of this type of uncertainty induces the population into cyclic dynamics, where the dominance of direct reciprocity increases with uncertainty. The model is also able to capture the inequality in contributions and the increase in polarization observed in the experiments (see Figure 4b).

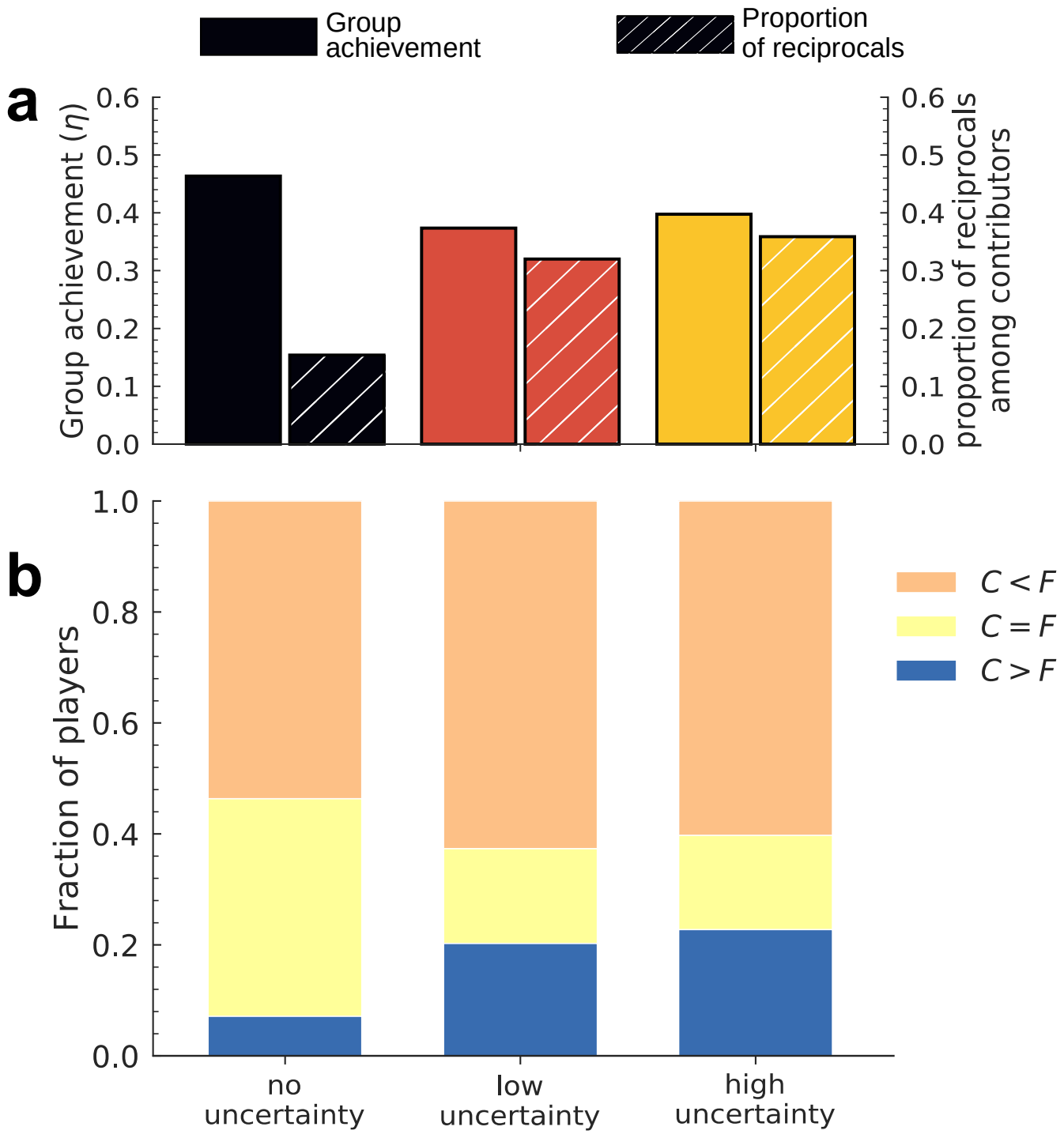

**Figure 4. Emergence of reciprocal strategy and polarization in a stochastic evolutionary model.** Panel (a) shows how that group achievement does not vary significantly with timing uncertainty (left y axis), while the predominance of the reciprocal strategy over the strategies that contribute to reach the target increases (right y axis and boxes with slashes). Panel (b) shows that the polarization increases when there is timing uncertainty. These results reproduce the trend observed in the experimental data. ($\boldsymbol{Z = 50, N = 6, r = 0.9, E = 40, \tau = 120, \beta = 0.003}$, see Methods for a detailed explanation of the parameters).

## Discussion

Despite the simplicity of our experimental setup and associated theoretical model, the trends identified may hold potential lessons for dealing with uncertainty in local and global governance. Our results show that contrary to the outcome for other types of uncertainty, timing uncertainty promotes early investments by the participants. Moreover, timing uncertainty appears to increase polarized reactions among the participants, suggesting a co-existence of risk-prone and risk-averse preferences, while diminishing the number of players contributing a fair share and increasing those that give less or more. Interestingly, our result relates nicely with recent findings in the context of behavioral dynamics in urban settings. It has been shown that uncertainty associated with big cities intensified both risk-seeking and risk-taking reactions, while the predictability of small villages encouraged more homogeneous and intermediate choices[40]. Such heterogeneity highlights the intricacy of the study of the emergence of polarized preferences beyond contagious processes[41], with potential implications in various socio-political and ecological contexts.

At the same time, conditional behaviors emerge in the presence of timing uncertainty, and groups that were able to coordinate through reciprocal behavior were more successful than those that simply played a fixed strategy or compensated for those giving not enough. This heavily contrasts with the predominant behavior when the length of the game is certain, in which most players unconditionally opt for a fair share *F*. Our results suggest that, when the future is uncertain and stakeholders are aware of it, they tend to respond early, encouraging others to reach the target together. This effect may be potentially maximized when combined with communication, institutions or costly commitments [20,42–45].

**Acknowledgements**

E.F.D. is supported a F.W.O. (Fonds Wetenschappelijk Onderzoek) Strategic Basic Research (SB) PhD grant and J.G. is supported by an F.WO postdoctoral grant. This work was also partially supported by the F,W.O. project with grant nr. G.0391.13N, the FuturICT2.0 (www.futurict2.eu) project funded by the FLAG-ERA JCT 2016, and by FCT-Portugal (grants PTDC/EEI-SII/ 5081/2014, PTDC/MAT/STA/3358/2014, and UID/CEC/50021/2013).

**Author contributions**

E.F.D., J.G., G.K., J.C.B., F.C.S. and T.L. conceived the experiments. E.F.D. and J.G. performed the experiments. E.F.D., J.G. and T.L. analyzed the data. All authors evaluated the data analysis. E.F.D., F.C.S. and T.L conceived the model. All authors wrote and approved the paper.

**Data availability.**

The datasets generated and analyzed during the current study are available from the corresponding author on reasonable request.

**Code availability**

The code of the experimental framework developed to perform the experiments is available at https://github.com/Socrats/beelbe.

# Methods

**Experiments.** The results of our experiment were obtained by testing 246 participants (41% females) that were divided into 41 groups of six subjects each in a computerized experiment (using the software available at https://github.com/Socrats/beelbe). Ethical approval was obtained from the Ethical Commission for Human Sciences at the Vrije Universiteit Brussel to perform this experiment. Most of the participants were bachelor/master students of either the Université Libre de Bruxelles or the Vrije Universiteit Brussel. The average age of participants was 24 (with a standard deviation of ~4 years). During each session of the experiments, participants were assigned randomly into groups and were not allowed to communicate (physical barriers were set up between them). Participants never knew who the other members of their group were.

In the control treatment (NU), 12 groups played the collective-risk dilemma (see main text) defined as in[8] with the difference that the game was not framed as a climate change scenario, which makes the results more general to other scenarios of collective-risk where there is an uncertain deadline. Indeed, this type of N-person dilemma, combining non-linear and uncertain returns which are only reached in the future, are recurrent in many human collective endeavors, from public health measures to group hunting. In the treatment with low uncertainty (LU) and high uncertainty (HU), another 14 and 15 groups, respectively, played the variant of the game in which the final round was decided by a random process. After a minimum number of rounds (8 rounds in LU, and 6 in HU), the probability of the game ending after each round was $w$=1/3 and $w$=1/5 in LU and HU, respectively. To implement this uncertainty in LU (HU) a 6 (10) faces dice was thrown at the end of round 8 (6), and the game would continue if the result was higher than 2, thus generating the probability for ending the game in LU (HU).

**Identifying polarization of donations.** We divided the participants on our experiment based on their total donations throughout the game and their relationship to what we call *fair donation* $F$. This value corresponds to the minimum donation required for a group to be successful, if all participants invest the same, i.e., if all participants invest exactly $F$ the group will be successful. This value corresponds to half of the endowment ($F = E/2$). Therefore, we quantify the fraction of participants that contribute, in total, less ($C < F$), equal ($C = F$) or more ($C > F$) than $F$. In Figure 2, we show that the fraction of participants that invest $C < F$ and $C > F$ grows with *timing uncertainty*, while *fair* players ($C = F$) diminish. We associate this divergence of donations to an increase of *polarized* reactions.

**Identifying conditional behavior.** Conditional behaviors were assessed through the analysis of the average donation of each player as a function of the donations of the other group members in the previous round (see Figure 3). For this reason, the plot starts with the data after the first round. Also, we only take into account the data of the experiment before the target is reached, i.e., when the public account contains less than 120 EMUs. We adopt a weighted linear regression so that samples with smaller errors were more important than those with large ones. The weight of a point $i$ is calculated as $weight_i = {\sigma_{\bar{x}}}^{-1}$, where the errors $\sigma_{\bar{x}}$ were computed as :

$$\sigma_{\bar{x}} = \begin{cases} \infty, |x| = 1 \\ \sqrt{\frac{\sigma_{actions}}{|x|}}, |x| > 1 \end{cases}$$

Here $\sigma_{actions} = 4$, representing the range of the values an action can take, and $|x|$ indicates the number of samples used to calculate the average of the samples vector $x$. This way, points calculated from only 1 sample, almost do not count for the regression.

**Statistics.** In Figure 1a the averages and error bars (95% confidence interval) are computed across groups (n=12 for NU, n=14 for LU and n=15 for HU). For LU and HU, after the minimum number of rounds, $m_0$, n decreases, since some of the groups finished the game. For LU, $n = 9$ for round 9, $n = 6$ for round 10, $n = 4$ for round 11, $n = 2$ until round 17. For HU, $n = 12$ for round 7, $n = 7$ for round 8, $n = 4$ for round 9, $n = 2$ until round 13, and $n = 1$ until round 19. In Figure 1b each box plot display the proportion of groups that were successful for each of the treatments. 8 out of 12 groups were successful in NU, 9 out of 14 for LU and 7 out of 15 for HU. We performed a Chi-square test of independence ($P = 0.49952, n = 41, df = 2, \chi^2 = 1.38$) that issues that the differences between the group achievements among the treatments are not significant. The error bars in Figure 1c represent the 95% confidence interval. This figure shows the average round in which successful groups reached the target (n=8 for NU, n=9 for LU and n=7 for HU). In Figure 2 we calculate the fraction of successful players for each treatment that assume one of 3 contribution behaviors. The total number of players in each of the fractions ($C < F$, $C = F$, $C > F$) is (9, 21, 18) for NU, (13, 17, 24) for LU and (13, 8, 21) for HU. The error bars in Figure 3 were calculated as described in the "identifying conditional behavior" section (see SI for detail on the number of samples for the correlation analysis).

**Game theoretical model.** As an alternative to considering fully-rational agents, we describe the behavioral dynamics through an evolutionary process[46], in which individuals tend to copy those appearing to be more successful. We consider a finite population of *Z* individuals, who interact in groups of size *N*, in which they engage in the collective-risk dilemma with multiple rounds. Each individual can adopt one of the $n_s$=5 strategies that mimic the behaviors observed on the experimental data: *always-2*, *always-4*, *always-0*, *compensator*, and *reciprocal*. The first three strategies are unconditional, i.e., they will always invest the same, independently of the behavior of the other group members. Differently, compensator and reciprocal are conditional strategies that adapt their behavior to the rest of the group according to a threshold of total investments per round. We consider this threshold to be 10, which is exactly half of the maximum investment per round, without the focal player. In this manner, compensators always start investing 2, and, afterwards, invest 0 as long as the sum of contributions of the rest of group members in the previous round is above 10 units; otherwise they will invest 4 EMUs. The behavior of *reciprocal* is the exact opposite of *compensators*: They start by investing 2 EMUs and afterwards they invest 4 EMUs as long the sum of contributions of the other members of the group in the previous rounds is above or equal to 10 units; otherwise they invest 0. We do not assume any population structure, such that individuals are equally likely to interact with each other (the so-called well-mixed assumption). The success (or fitness) of an individual can be computed as the average payoff obtained from playing in multiple groups randomly sampled from the population. As a result, all individuals adopting one of the $n_s$=5 strategies can be seen as equivalent, on average.

To study the behavioral resulting from this set of strategies, we adopt a stochastic birth-death process combined with the pairwise comparison rule[39] to describe the social learning dynamics of each of the strategies in a finite population. At each time-step, a randomly chosen individual $A$ has the opportunity to revise their strategy by imitating (or not) the strategy of a randomly selected member of the population $B$. The imitation will occur with a probability which increases with the fitness difference between $A$ and $B$. Here we adopt the Fermi function $p \equiv \left[1 + e^{\beta(f_A - f_B)}\right]^{-1}$, where $\beta$ controls the intensity of selection (we use $\beta = 0.003$), and $f_A$ ($f_B$) is the average fitness of $A$ ($B$). In addition, we consider that, with a mutation probability $\mu$, individuals adopt a randomly chosen strategy, freely exploring the strategy space. Overall this adaptive process defines a large-scale Markov process, whose complete characterization becomes unfeasible as one increases the population size and number of strategies[47]. However, this analysis of this stochastic dynamics is largely simplified in the limit of rare mutations. In this case, we are able to compute analytically the relative prevalence of each of the different strategies. Moreover, as shown in the Supporting Information by means of large-scale computer simulations, this analytical approximation turns out to be valid for much wider interval of mutation regimes. In this limit, when a new strategy appears through mutation, one of two outcomes occurs long before the occurrence of a new mutation: either the population faces the fixation of newly introduced strategy, or the mutant strategy is wiped out from the population. Hence, there will be a maximum of two strategies present simultaneously in the population[48,49]. This allows one to describe the behavioral dynamics in terms of a reduced Markov Chain of size $n_s = 5$, whose transitions are defined by the fixation probabilities $\rho_{ij}$ of a single mutant with strategy $j$ in a population of individuals adopting another strategy $i$. This probability is given by $\rho_{ij} = \left(1 + \sum_{m=1}^{N-1} \prod_{k=1}^{m} \frac{T^-(k)}{T^+(k)}\right)^{-1}$ [39,50,51], where $T^-(k)$ ($T^+(k)$) is the probability to decrease (increase) the number of individuals with the mutant strategy: $T^{\pm}(k) = \frac{k}{Z}\frac{Z-k}{Z}\left[1 + e^{\mp\beta[f_i - f_j]}\right]^{-1}$. In the limit of neutral selection ($\beta = 0$), the fixation probabilities become independent of the fitness values and equal to 1/$Z$, offering a convenient reference scenario (see below). Since we will have at most two different strategies in the population, we can calculate the fitness $f_a$ of a strategy $a$, in a finite population of size $Z$ and $k$ individuals of strategy $a$ and $Z - k$ of strategy $b$, as $f_a = \binom{Z-1}{N-1}^{-1} \sum_{k=0}^{N-1} \binom{k-1}{j} \binom{Z-k}{N-j-1} \Pi_{ab}(k+1)$ [27], where $\binom{Z-1}{N-1}^{-1} \sum_{k=0}^{N-1} \binom{k-1}{j} \binom{Z-k}{N-j-1}$ represents a hypergeometric sampling (sampling without replacement) of the population and $\Pi_{ab}(k+1)$ is the payoff of strategy $a$ when facing strategy $b$, while the group is composed of $k+1$ individuals with strategy $a$. We numerically estimate pairwise payoffs $\Pi_{ij}$ between every strategy pair $i$ and $j$, for each possible composition of a group of $N$ participants with $k$ members of using strategy $i$ and $N - k$ using strategy $j$. This is achieved by averaging over $10^3$ games for each composition of the group and treatments (NU, LU and HU, using the same parameters adopted in the lab experiments).

Equipped with these tools, one can now compute the prevalence of each strategy through the stationary distribution $p_w$ of the $n_s$-states Markov chain. The stationary distribution characterizes the average time the population spends in each monomorphic state $w$. This can be computed analytically as the normalized eigenvector associated with the eigenvalue 1 of the transposed of the transition matrix $M$ of the Markov Chain[48,49,51] with all fixation probabilities $\rho_{ij}$. To compute the expected group achievement ($\eta$) shown in Figure 4a, we weight the probability of success of each of the monomorphic states (populations with only one of the 5 strategies) by their predominance (given by the stationary distribution), i.e., $\eta = \sum_w p_w H_w$, where $p$ is a row vector containing the stationary distribution, and H is a column vector containing the probability of success of each monomorphic state. In our case, the probability of success is always 1 for populations of *always-2*, *reciprocal* and *always-4* players, while it is 0 for *always-0*. Compensators are only successful if the game lasts more than 10 rounds, therefore their probability of success is $(1-w)^{10-m_0}$, since the random process that decides the final round under *timing uncertainty* follows a geometric distribution. The calculation of the fraction of players that contribute less, equal or more than $F$, used in Figure 4b, is done in a similar fashion. In this case, we need to calculate the probability that a population consisting of each of the monomorphic states will contribute $C < F$, $C = F$, $C > F$. Since we can calculate the contributions of the players depending on the number of rounds of the game, the computation of these probabilities is straightforward. The *always-2* players will always invest $F$, *always-4* and *reciprocal* will contribute $C > F$, and *always-0* $C < F$. Once more, compensators will only contribute $C > F$ if the number of rounds is bigger than 10, otherwise they contribute $C < F$. We then multiply these probabilities by the stationary distribution to obtain the fraction of players that assume each of the previous behaviors. In Figure 4 all other parameters controlling the environment are set to be the same as in our lab experiments: the risk $r = 0.9$, the initial endowment $E = 40$ and the target is $\tau = 120$. To obtain further intuition behind the emergence of cooperation, group achievement and strategies in each treatment, we also analyze the typical flow of probability between the different monomorphic states in the presence and absence of uncertainty (see invasion diagrams in Figure S6). Arrows represent transitions favored by natural selection, i.e., those whose fixation probability exceeds 1/Z (associated with the fixation probability of a mutant under neutral evolution). A strategy for which no mutant adopting any other strategy has a selective advantage is said to be an *evolutionary robust strategy* (ERS)[52], a convenient measure of strategic stability in this context. As shown in Figure S6, timing uncertainty changes the set of evolutionary robust strategies, benefiting the flow of probabilities towards the adoption of reciprocal strategies.

# Supporting information for

# Timing uncertainty in collective risk dilemmas encourages group reciprocation and polarization

Elias Fernández Domingos, Jelena Grujić, Juan C. Burguillo, Georg Kirchsteiger, Francisco C. Santos, and Tom Lenaerts

## Nomenclature

For the remainder of this document we will refer to each one of the 3 treatments analyzed in this manuscript in the following way:

- Treatment 1: no uncertainty treatment – NU
- Treatment 2: low uncertainty treatment – LU
- Treatment 3: high uncertainty treatment - HU

## Extended Methods

### ***Experimental procedure***

During each session of the experiment, all participants were required to read the instructions on the screen of their assigned computer before the start. The same instructions were provided on a printed copy that they could consult throughout the experiment. After reading, all participants went through a test with the goal to check their understanding. In case of problems the coordinators discussed with them their errors ensuring that everything was clear. Throughout the experiment each participant observed on his or her screen the amount left in his or her *private account* and the actions of each of the group members in the previous round. They are thus able to keep track of the behavior of their group mates, but do not know their identity. They could not observe the current state of the *public account*, yet, they were encouraged to keep track of it by asking them how much they believe is in the account at each round. After the final round (in the treatment with uncertainty they could observe the result of the random value produced by the dice that defined the end of the game), the participants could see on the screen how much was invested in the *public account* in total and how much was left in their *private account*, as well as the conversion to euros. In the case this value was below the target, a message would show the result of the dice that decided whether they would lose or not the remaining endowment. Finally, before the participants were allowed to leave the laboratory, they were requested to complete a small survey about their experience during the experiment.

Our experiment models the effect of timing uncertainty on the collective-risk dilemma. Therefore, there is a stochastic component that we must explain to participants carefully. In order to do this, we used the known example of a virtual dice. For instance, to explain that there is a probability of 1/3 that the game would finish after round 8 in the low uncertainty (LU) treatment, we explain that the computer will "throw a virtual dice of 6 faces, and if the result is either 1 or 2, then the game will end". We also tell participants that, on average, the game takes 10 rounds, to give them an intuition about the distribution of this stochastic process. The details of the instructions for all three treatments can be found in the following sections.

***Instructions for the control treatment (no uncertainty – NU)***

Each participant had access to the following instructions, both in digital and paper format:

Instructions to the experiment

**Welcome to this experiment where you can earn money!**

You are about to participate in an experiment on iterative decision-making, conducted by researchers from the *Vrije Universiteit Brussel* and the *Université Libre de Bruxelles*. In this experiment, you will earn some money, and the amount will be determined by your choices and the choices of the other participants.

**Your privacy is guaranteed:** The other participants will not know who you are during the experiment and the results of the experiment are stored in an anonymous manner.

It is very important that you remain silent during the whole experiment, and that you never communicate with other participants, neither verbally, nor in any other way. When in doubt or when you have a question, please just raise your hand and an experimenter will approach you. If you do not remain silent, or if you behave in any way that could potentially disturb the experiment, you will be asked to leave the laboratory, and you will not be paid.

All your earnings during the experiment will be expressed in Experimental Monetary Units (EMUs), which will be transformed into Euros with a change rate of 0.75 Euro to 1 EMU. At the end of the experiment, a show up fee of 2.5 euros will be added to your earnings.

You will be paid privately by bank transfer to your account within a week after the experiment. At the end of the experiment you will be requested to provide your **IBAN number and BIC code** to make the transfer.

Before starting, you will be randomly assigned into a group. You will never know the identity of the other participants of the group. However, the experiment takes 10 rounds and you will be able to observe the actions of the previous round of every member of your group, starting from round 2.

Login to the experiment

Before the experiment can start, please, enter the user login and password you have been given into the login page displayed in the browser of the computer assigned to you.

Once you have logged in, you will be able to see on your screen the same instructions that are written on this paper.

**Wait for the instructor's signal before you proceed.**

General Information

At the beginning of the experiment you will be randomly assigned to a group, which will include 5 other randomly selected participants.

During the whole experiment, you will interact only with those 5 other group members.

At the beginning of the experiment you and each other group member will receive **a personal endowment of 40 EMUs**.

The whole experiment consists of 10 rounds of the following game.

In each round of the game, you have to decide whether to add 0, 2 or 4 EMUs in a **public account**.

If the public account contains at least **120 EMUs** after the $10^{th}$ round, **each member of your group will keep their savings**, i.e. the EMUs of your endowment that were **not** put in the public account.

However, **if this minimum is not reached,** the computer will "throw a virtual dice" and **each group member will lose his or her remaining EMUs with a 90% chance (9 times out of 10)**.

Thus, with a 10% chance (1 out of 10) you will keep the remaining EMUs in your private account.

Course of Action

Every round has the same structure and consists of the following **steps**:

Step 1: Choice of how much to contribute (0,2 or 4).

Step 2: Make a prediction about the amount in the public account.

When the experiment reaches its final round, you will move to the final 3rd step:

Step 3: Check if the threshold of the public account has been achieved and calculation of final payoffs

<u>Step 1: The contribution choice</u>

In the Step 1, every member will be asked **"How many EMUs do you want to contribute to the public account"**. Three buttons are provided: **0, 2 and 4 EMUs**. You can select the amount by clicking the button, as is shown in the figure below:

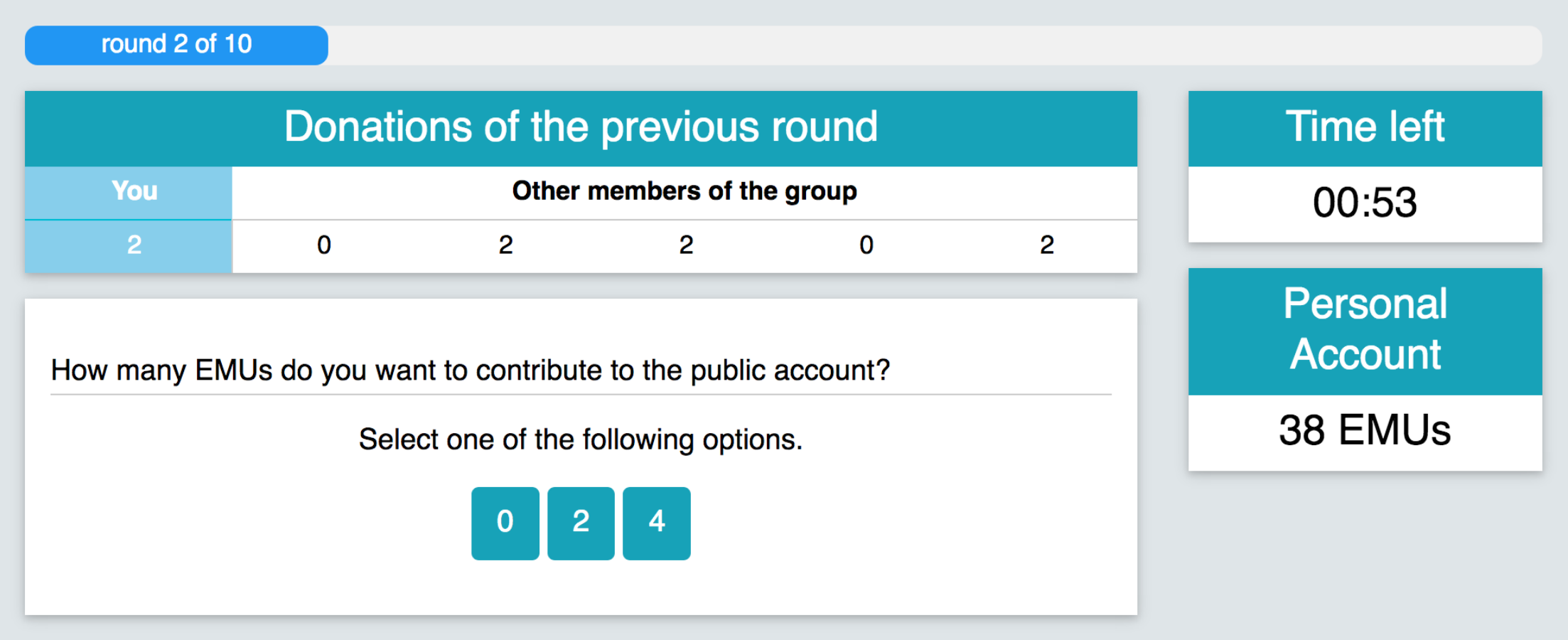

*Image 1: View of Step 1.*

On the right side of the screen you can see the time you have left to make your decision and the amount of EMUs in your **"Personal Account"**. **You must make your decision within** **the time displayed on the screen**. The "Time left" square will start blinking when you are getting out of time. Nothing happens when the time runs out, yet if you take too long to make a decision the experiment will take too long. Please respond as quickly as possible.

The table "Donations of the previous round" shows the values donated by all the members of **your group** in the previous round. In the first column, you see **your own donation** from the previous round. In the other columns, you see the decisions of the other users. The choice of each group member will always be shown in the same column. This information about the previous donations is only available after the first round.

Step 2: Predict the content of the public account

After step 1, you will go to a next screen. On this screen, you are asked the following question: "**Please, estimate the current total content of the public account**". You should enter an estimation of how many EMUs you think the **public account** contains **in total** after all members (including you) have made their donations in the current round.

This is an example of what you will see in this step:

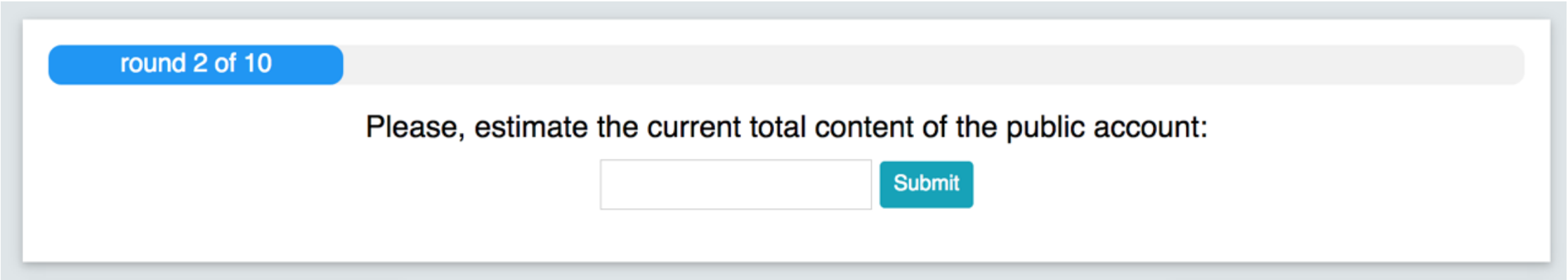

*Image 2: View of Step 2.*

Step 3: Last round and calculation of final payoffs

After the last round, you will jump to a final screen.

If the accumulated contributions to the account are equal or higher than **120** EMUs, then you will be informed that you can keep the amount of the endowment that you **did not** put in the public account.

For example, if you put in total 20 EMUs of your endowment (40 EMUs) in the public account during the experiment, you will gain the remaining **20** EMUs (i.e. 40 - 20). This amount is converted into Euro's.

The screen will show the following text: "*CONGRATULATIONS! Your group collected XXX EMUs, which is greater or equal to 120 EMUs. So **you may keep the amount remaining in*** ***your private account****. Please fill in the amount in Euro's you see on this screen on the payment document you received before clicking the continue button. This amount consists of both your private winnings and the show-up fee."*

However, **if the minimum of 120 EMUs is not reached,** the computer will "throw a virtual dice" and **all group members will lose all their remaining EMUs with a 90% chance.** There are thus two possible outcomes:

On one hand, **with a chance of 9 out of 10, the screen will show**: "*Your group collected XXX EMUs, which is lower than 120 EMUs. The server has generated a random number between 1 and 100. The resulting value is YYY, which is smaller than 91. This means that you all* ***lose the remaining endowment*** *in your private accounts. Please fill in the amount in Euro's you see on this screen on the payment document you received before clicking the continue button. This amount is the show-up fee.*"

On the other hand, **with a chance of 1 out of 10, the screen will show**: "*Your group collected XXX EMUs, which is lower than 120 EMUs. The server has generated a random number between 1 and 100. The resulting value is YYY, which is bigger than 90. This means that you all* ***win the remaining endowment*** *in your private accounts. Please fill in the amount in Euro's you see on this screen on the payment document you received before clicking the continue button. This amount consists of both your private winnings and the show-up fee.*"

<u>End of experiment questionnaire</u>

At the end of the experiment you will be directed to a form containing a short questionnaire. Please answer to all the questions honestly, the information you add here is an important part of this experiment. Any information that you may include in this form will remain completely anonymous and cannot be linked to you in any way. Once you have finished filling in the questionnaire, please, click the button ***submit***.

**At the end of the experiment, you will be called by one of the organisers to make the payment official. Please stay seated and do not talk until you are called and have left the room.**

<u>Please note:</u>

Communication is not allowed during the whole experiment. If you have a question, please raise your hand.

All decisions are made anonymously, i.e. no other participant learns the identity of the other decision makers.

The payment is also anonymous,  no participant learns from us about the amount that another participant received in the experiment.

***Instructions for the low uncertainty treatment (LU)***

The instructions of the treatment with low *timing uncertainty* differ slightly from that of the previous treatment. Below we describe only the sections that change with respect to the control treatment (NU).

**Instructions to the experiment**

Before starting, you will be randomly assigned into a group. You will never know the identity of the other participants of the group. However, the experiment takes **at least** 8 rounds and you will be able to observe the actions of the previous round of every member of your group, starting from round 2.

General Information

The whole experiment consists **of minimum 8 and on average 10 rounds, but it could be more than that.**

The probability that the next round will happen after round 8 is 2/3. This means that at the end of each round, starting from round 8, a **"virtual fair dice"** with 6 faces, will be thrown. If the result is **"1"** or **"2"**, the game will end. Otherwise, when the result is "**3**", "**4**", "**5**" or "**6**", the game will continue to the next round, in which the process is repeated.

If the public account contains at least **120 EMUs** after the final round, **each member of your group will keep their savings**, i.e. the EMUs of your endowment that were **not** put in the public account.

Course of Action

Every round has the same structure and consists of the following **steps**:

Step 1:  Choice of how much to contribute (0,2 or 4).

Step 2: Make a prediction about the amount in the public account.

The end of the experiment is decided by a random process. Starting from **round 8** the game will go through a 3rd step:

Step 3: Check if the experiment should end by throwing a **"virtual fair dice"**.

When the experiment reaches its final round, you will move to the final 4th step:

Step 4: Check if the threshold of the public account has been achieved and calculation of final payoffs

Step 3: Check if the game should end
The probability that the next round will happen after round 8 is 2/3. This means that at the end of each round, starting from round 8, a **"virtual fair dice"** with 6 faces, will be thrown. If the result is **"1"** or **"2"**, the game will end. Otherwise, when the result is "**3**", "**4**", "**5**" or "**6**", the game, will continue to the next round, in which the process is repeated. This means that the experiment will have a minimum of 8 rounds.

There is a **"virtual fair dice"** for each group. Thus, the experiment can have different rounds, depending on which group you are in.

You will be able to see on the screen the following text if the game continues: "*The result of the dice was X, which is different from "1" or "2". Therefore, **the experiment will continue** to the next round. Please, click now on the button **Ok**.*"

However, if the result of the dice is "1" or "2", the screen will show the following text: "*The result of the dice was X. Therefore, **the experiment will end**. Please, click now on the button **Ok**.*"

***Instructions the high uncertainty treatment (HU)***

The instructions for the treatment with high *timing uncertainty* differ from treatment 2 (low timing uncertainty) only in that the minimum number of rounds is 6 and the probability that the game ends afterwards is 1/5 (or 2/10). However, the average number of rounds remains 10. Below we describe only the sections that differ from the LU treatment:

General Information

The whole experiment consists **of minimum 6 and on average 10 rounds, but it could be more than that.**

The probability that the next round will happen after round 6 is 8/10. This means that at the end of each round, starting from round 6, a "virtual fair dice" with 10 faces, will be thrown. If the result is "**1**" or "**2**", the game will end. Otherwise, when the result is "**3**", "**4**", "**5**", "**6**", "**7**", "**8**", "**9**" or "**10**", the game will continue to the next round, in which the process is repeated.

Step 3: Check if the game should end

The probability that the next round will happen after round 6 is 8/10. This means that at the end of each round, starting from round 6, a **"virtual fair dice"** with 10 faces, will be thrown. If the result is "**1**" or "**2**", the game will end. Otherwise, when the result is "**3**", "**4**", "**5**", "**6**", "**7**", "**8**", "**9**" or "**10**", the game will continue to the next round, in which the process is repeated. This means that the experiment will have a minimum of 6 rounds.

***Testing participant's understanding***

After reading the instruction, participants are requested to complete a short questionnaire that tests their understanding. Participants are not allowed to start the experiment until they answer to all questions correctly.

**Supplementary experimental results**

In Figure 1 of the main text we display the averaged cumulative contributions per group for each of the 3 treatments of our experiment. In Figure S1 we present the average contributions per round. Here we observe that, for the successful groups (met target = True), the investments in the treatments with uncertainty were always higher than in the control treatment (NU) before the minimum number of rounds. The figure also shows that for NU, the difference between successful groups and non-successful ones is mostly related to the contributions in the last two rounds. While the successful groups increase their contributions slightly by the end (compensate for other participants), the non-successful groups lower them. This highlights the importance of coordination when the time is certain. For the treatments with *timing uncertainty* this effect disappears, and successful groups contribute more and earlier.

To reveal the differences between the treatments, we used a t-SNE 2D visualization (Figure S2) of the multidimensional vector that represents each participant (which is represented by a tuple of the actions she took at each round and the total donations of their group mates in the previous round). This representation is able to cluster (with some noise) the participants of each treatment, which allows us to confirm visually that there are differences between them. Each color in the figure represents one treatment and the colors are the same as those in Figure 1 of the main manuscript (NU black, LU red, HU yellow).

In Figure S3 we show a comparison between the participants behaviors in the first and second parts of the game for the participants that met the target (this corresponds to Figure 2 of the main manuscript) and those that didn't. We can observe an increasing difference between the *generous* players (those that invest $C > F/2$) in the first half of the game as *timing uncertainty* grows. This highlights the importance of early investments to nudge participants into cooperation when there is a *shadow on the future*.

Figure S4 shows the frequency of each of the 3 possible actions, depending on the group investments (without the focal player) in the previous round. This figure combines the participants from both the successful (met the target) and unsuccessful sessions. While in the treatment without *timing uncertainty*, action 2 is predominant, in the treatment with uncertainty, we observe and increase of the extreme actions (0 and 4). Moreover, action 4 occurs with higher frequency when the previous donations of the other members of the group were higher than 10, and action 0 when they were lower.

We have also extended our analysis of the conditional behavior to study how it depends on time (first half and second half of the game) and on the previous action of the focal player. Yet, these results must be analyzed with care, since the amount of data used to calculate each point, becomes smaller with each new feature taken into account (which is reflected in the error bars of the figures). Figure S5 indicates that there are differences in conditional behavior depending on the half of the game. This is expected, since Figure 2 of the main text already shows that in the treatments with *timing uncertainty*, the investments are higher in the first half.

**Error! Reference source not found.** shows the correlation results and the associated p-values for the results presented in Figure 3 of the main text. The correlations for the treatment without uncertainty are negative, which indicates a compensatory behavior. For the treatments with uncertainty, however, these correlations are positive, indicating a reciprocal or Tit-for-Tat behavior. Nevertheless, only the correlations for the successful players of LU and HU are statistically significant ($P < 0.001$).

**Dependency analysis**

The results in Table S2-S4 are all obtained using an ordered logistic regression or *cumulative link model*[1], implemented in the *polr* function implemented in the MASS package in R. We estimate the probability of taking an action (0, 2 and 4) depending on a series of features. This allows us to study how the actions of the participants on the experiment depend on these features. We used this analysis to select the most relevant features for the behavioral representation of a participant, which was then used to produce the results of Figure 2 and 3 of the main manuscript. The *polr* function differs from a multimodal regression in that it performs a ordered logistic regression, i.e., it takes into account the order of the labels. This is important in our case, since the donations 0, 2 or 4 are ordered.

The features included in the regression are:

- public_account: The public account of the game, i.e., the cumulative sum of investments of all participants
- private_account: The private account of the participant, i.e. the remaining endowment at a given round.
- round_donations_others: The donations of the members of the group in the previous round, without the focal player.
- actions_prev: The action of the focal player in the previous round.
- GameHalf2ndHalf: The half of the game in which the action takes place (1$^{st}$ half, 2$^{nd}$ half).

- rnd: A random binary number (0 or 1). This is used to check that the regression is producing correct results. The actions of the players should not depend on a random variable.
- public_account:GameHalf2ndHalf: Interaction term between the public account and the game half. Represents the degree to which there is an interaction between these two variables.
- actions_prev:GameHalf2ndHalf: Interaction term between the previous action of the player and the game half.
- private_account:GameHalf2ndHalf: Interaction term between the private account and the game half.

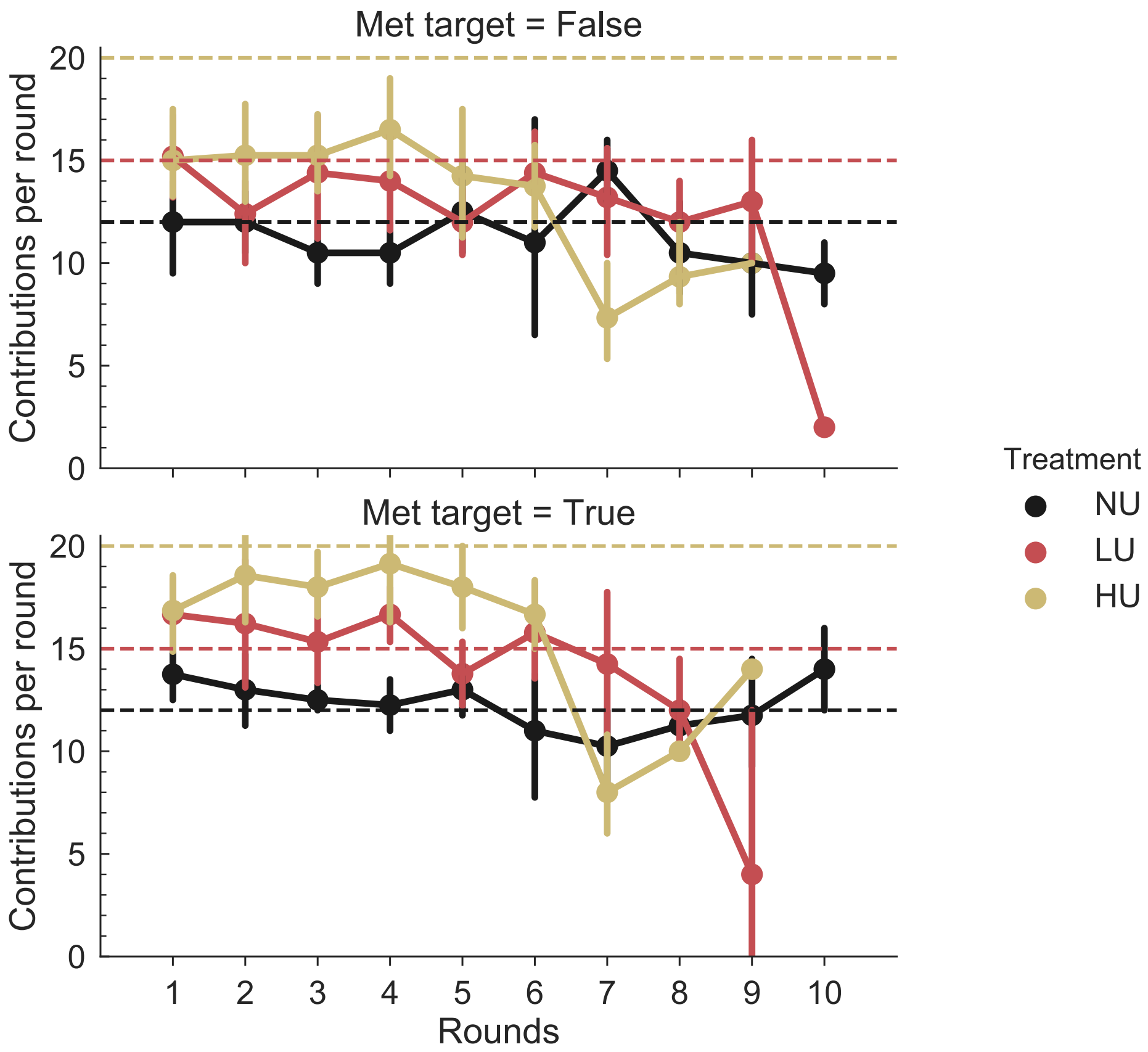

Figure S1. Average joint contribution per round and per group. The plots are separated by whether the group reached or not the target collective investments. The dashed lines show *the* fair sum of contributions per round if the game had 10 rounds for each of the treatments (black – NU, red – LU, yellow – HU). We only show the contributions before the target is reached.

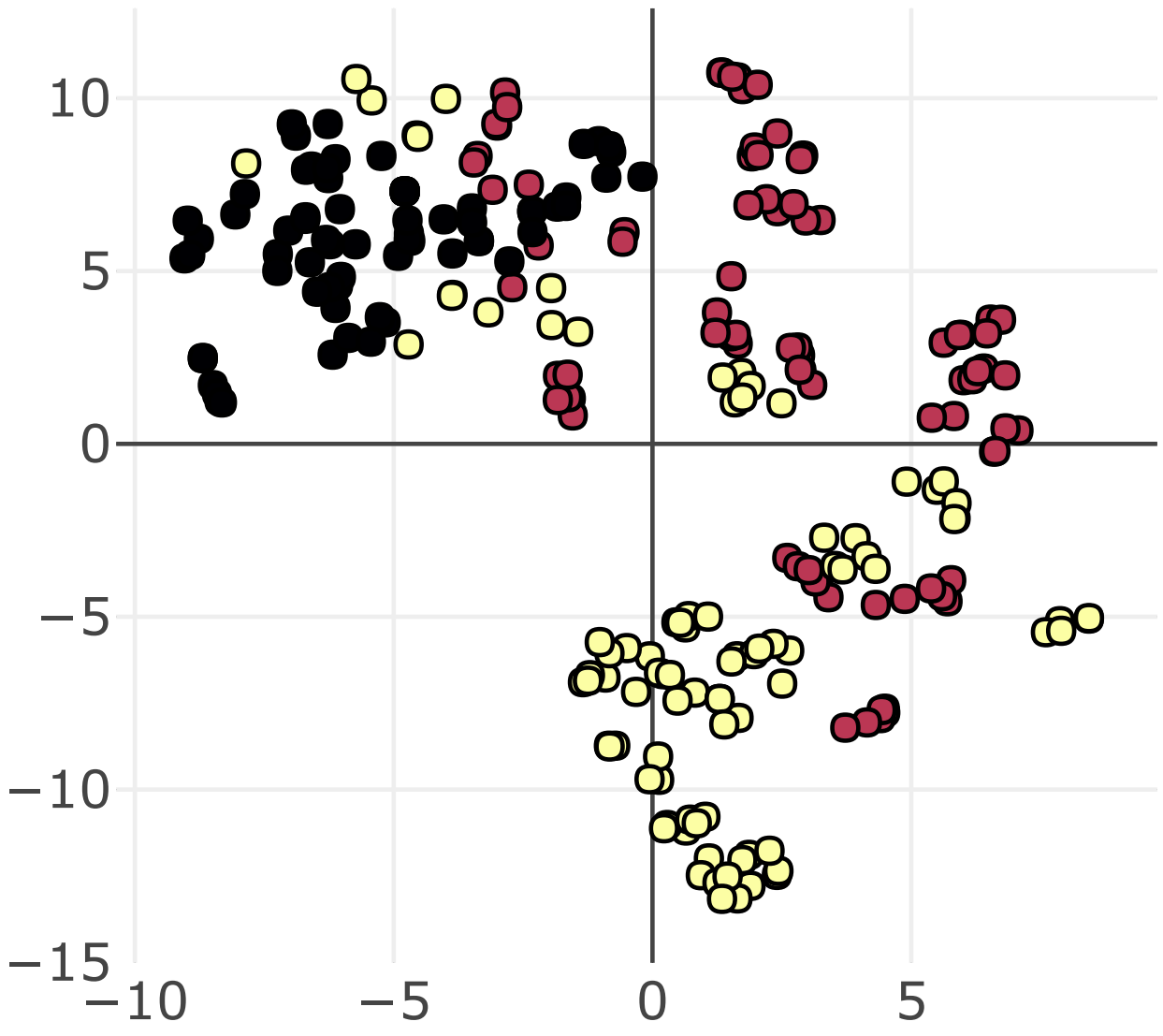

Figure S2. T-SNE representation of participants' behaviors. The t-SNE plot shows a 2-dimensional representation of the behavior of each player based on decisions taken by the individuals and their group mates: Similar behaviors are positioned in nearby points and dissimilar ones are characterized by distant points. We can observe that the data points of each treatment form clusters, which suggests that the behavior of participants is different in each treatment (the axis dimensions represent the 2D projection of the features used in the t-SNE embedding).

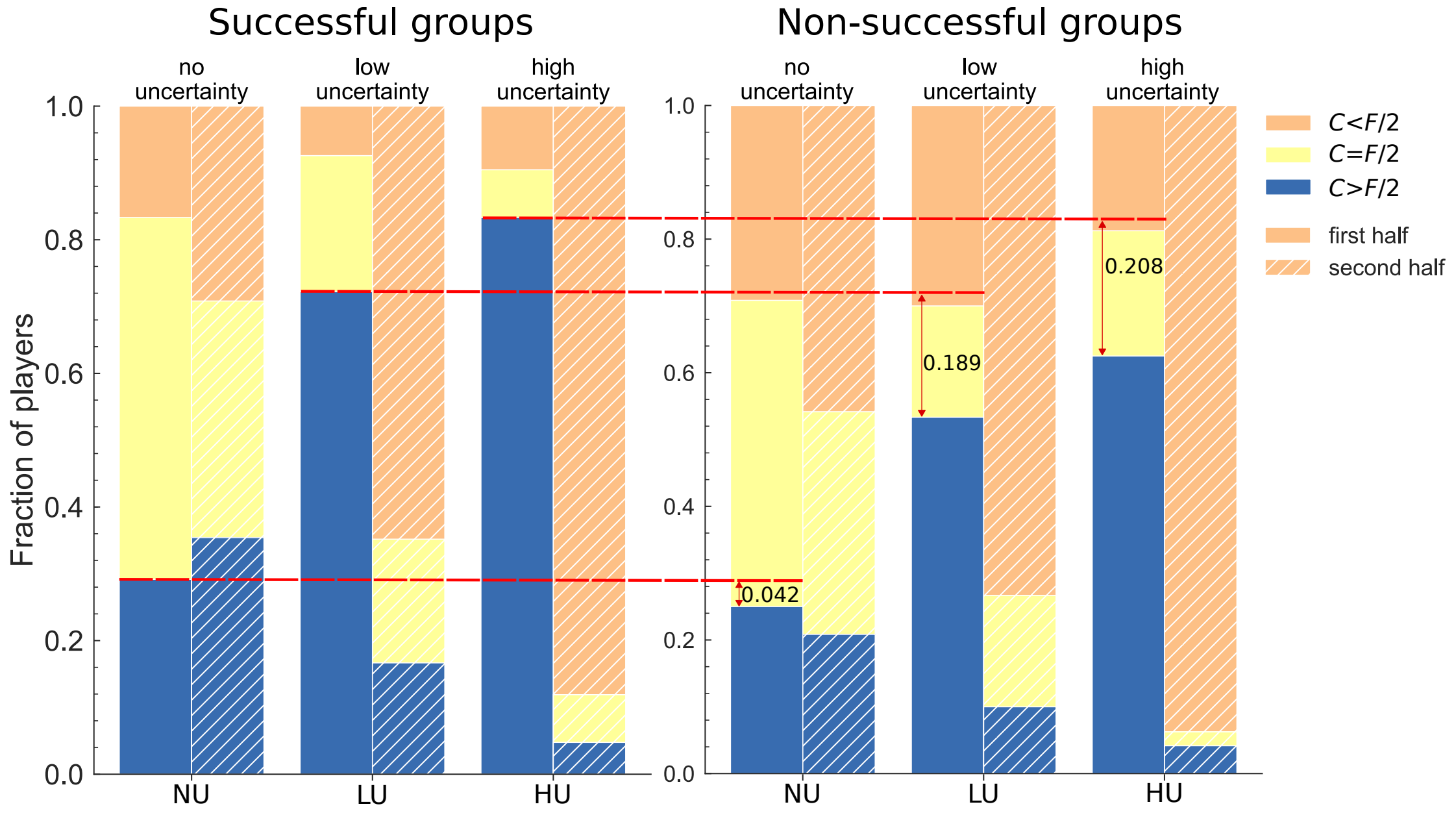

Figure S3. Distribution of players according to their investments. The plots are separated by whether the groups reached (successful) or not (non-successful) the target. The plots show the fraction of players that invested more, equal or less than $\boldsymbol{F/2}$, i.e., half of the *fair donation*, in the first and second half of the game. If every player invested in total $\boldsymbol{F}$. during the game, the group would reach the target with exactly 120 EMUs. For T2 and T3 we consider half of the game to be $\boldsymbol{m_0/2}$. We can observe that the number of participants that invest more than $\boldsymbol{F/2}$ in the first half of the game (non-procrastinators), increase considerably in the treatments with uncertainty. It is also noticeable that the difference in the fraction of non-procrastinators between the groups that met and did not meet the target, increases with uncertainty.

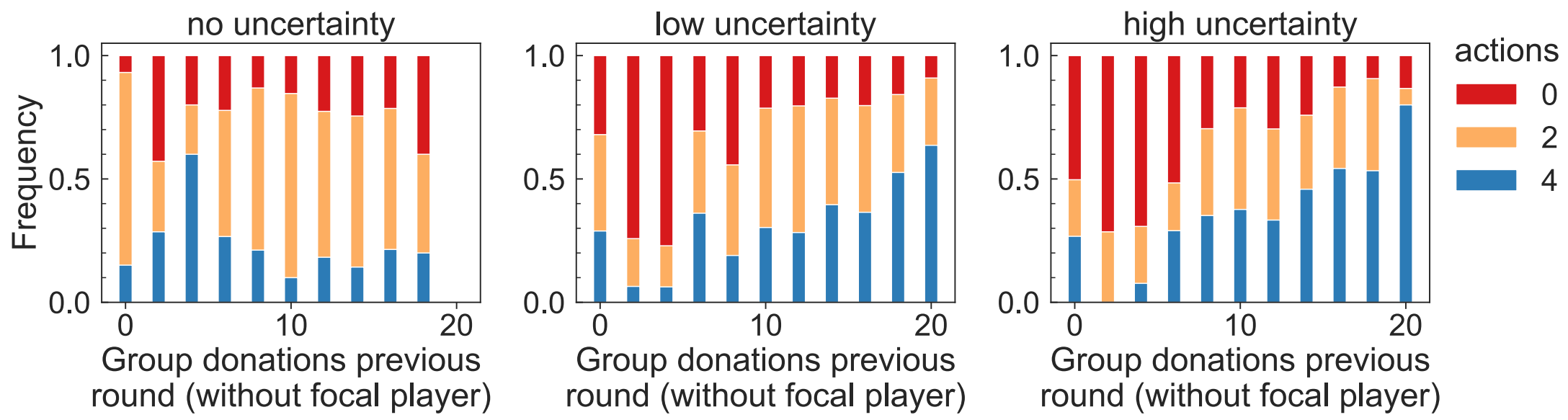

Figure S4. Conditional behavior per treatment. Distribution of actions in function of the donations of the group members (without the focal player) in the previous round. In NU (on the left) players display an almost uniform behavior over the set of donations of their group members in the previous round (2 is the predominant action),

with an increase of action 4 when the group members donated a total of 4, related to the players trying to compensate for their teammates on the last rounds. In contrast, in the treatments with timing uncertainty (center and right) players are more likely to donate 4 when their group members donated over 10 EMUs in the previous round and increase their 0 donations when the group members do the same, which corresponds to a Tit-for-Tat behavior.

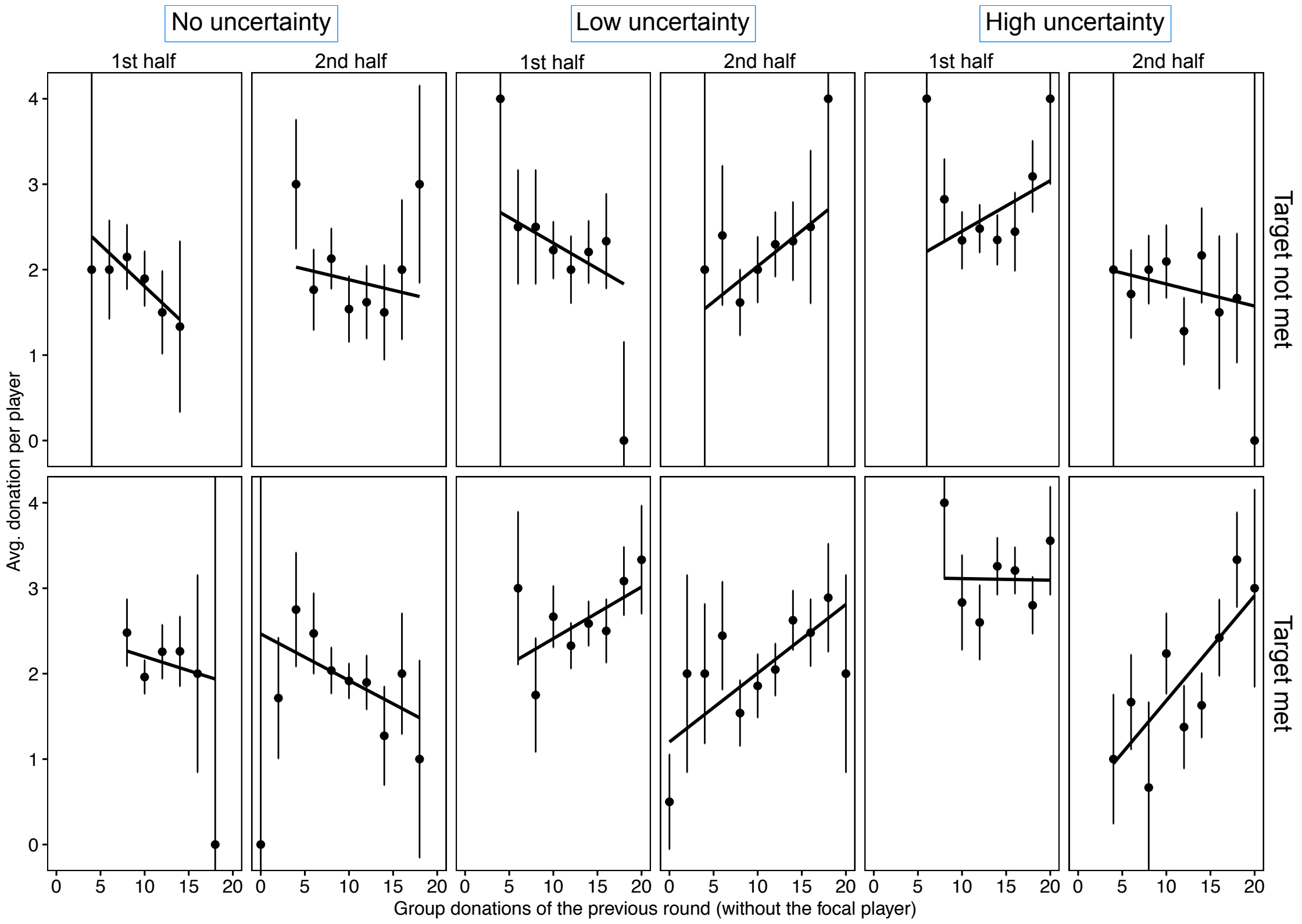

Figure S5. Analysis of conditional behavior separated by half of the game. Correlation between behavior of others and the behavior of a focal player separated both by treatment and by the first 5 rounds and the rest of the game.

| treatment | target | correlation | p-value | Number of observations |
|---|---|---|---|---|
| NU (No uncertainty) | FALSE | -0.117 | 0.086 | 216 |
| | TRUE | -0.076 | 0.114 | 432 |
| LU (Low uncertainty) | FALSE | 0.019 | 0.775 | 228 |
| | TRUE | 0.279 | < 0.001 | 408 |
| HU (High uncertainty) | FALSE | 0.092 | 0.114 | 300 |
| | TRUE | 0.268 | < 0.001 | 282 |

Table S1. Correlation between behavior of others and the behavior of a focal player. This table shows the Pearson correlation coefficients and the associated p-value between the sum of donations of the other members of a group in a previous round and the action of the focal player on the current round. These results are associated with Figure 3 of the main text. The correlation is positive ($P < 0.001$) for the successful players (target = TRUE) on LU and HU, while the ones that failed to reach the target have a correlation close to 0. The players in NU have a small negative correlation that indicates the presence of *compensating* behaviors.

***No uncertainty treatment (NU):***

| Coefficients | Value | Std. Error | t value | p value |
|---|---|---|---|---|
| public_account | -0.0370 | 0.0181 | -2.0436 | 0.0410 |
| private_account | -0.2503 | 0.0969 | -2.5822 | 0.0098 |
| round_donations_others | 0.0656 | 0.0650 | 1.0091 | 0.3129 |
| actions_prev | 0.3860 | 0.1749 | 2.2076 | 0.0273 |
| GameHalf2ndHalf | -13.4311 | 4.2979 | -3.1251 | 0.0018 |
| rnd | 0.0510 | 0.1660 | 0.3074 | 0.7585 |
| public_account:GameHalf2ndHalf | 0.0583 | 0.0204 | 2.8645 | 0.0042 |
| round_donations_others:GameHalf2ndHalf | -0.2008 | 0.0765 | -2.6266 | 0.0086 |
| actions_prev:GameHalf2ndHalf | 0.0642 | 0.2009 | 0.3194 | 0.7494 |
| private_account:GameHalf2ndHalf | 0.3827 | 0.1042 | 3.6719 | 0.0002 |

| Intercepts | | | | |
|---|---|---|---|---|
| 0\|2 | -10.1362 | 3.9585 | -2.5606 | 0.0104 |
| 2\|4 | -6.7677 | 3.9472 | -1.7146 | 0.0864 |

| | |
|---|---|
| Residual Deviance: | 1084.3870 |
| AIC: | 1108.3870 |

Table S2. Ordered logistic regression for the treatment without uncertainty. This table shows how much each feature influences the selection of an action for a player. The p-value threshold selected here is $p<0.05$. The results highlight that, among others, time (GameHalf) is a relevant feature, however, in the donation of the other members of the group (round_donations_others) does not influence significantly the selection of an action.

***Low uncertainty treatment (LU)***

| Coefficients | Value | Std. Error | t value | p value |
|---|---|---|---|---|
| public_account | -0.0556 | 0.0090 | -6.1725 | < 0.0001 |
| private_account | -0.4054 | 0.0199 | -20.3623 | < 0.0001 |
| round_donations_others | 0.1178 | 0.0405 | 2.9103 | 0.0036 |
| actions_prev | -0.1736 | 0.0955 | -1.8177 | 0.0691 |
| GameHalf2ndHalf | -11.4270 | 0.5859 | -19.5022 | < 0.0001 |
| rnd | 0.0986 | 0.1436 | 0.6867 | 0.4923 |
| public_account:GameHalf2ndHalf | 0.0130 | 0.0095 | 1.3606 | 0.1737 |
| round_donations_others:GameHalf2ndHalf | -0.0140 | 0.0471 | -0.2961 | 0.7672 |
| actions_prev:GameHalf2ndHalf | 0.3034 | 0.1132 | 2.6803 | 0.0074 |
| private_account:GameHalf2ndHalf | 0.3342 | 0.0199 | 16.7546 | < 0.0001 |

| Intercepts | | | | |
|---|---|---|---|---|
| 0\|2 | -16.4973 | 0.6035 | -27.3366 | < 0.0001 |
| 2\|4 | -14.3607 | 0.5973 | -24.0445 | < 0.0001 |

| | |
|---|---|
| Residual Deviance: | 1419.8400 |
| AIC: | 1443.8400 |

Table S3. Ordered logistic regression for the treatment with low uncertainty. The results for low uncertainty show that, like for no uncertainty, time is important. However, in this case, the donations of the other participants (round_donations_others) are also significantly affecting the action selection.

***High uncertainty treatment (HU)***

| Coefficients | Value | Std. Error | t value | p value |
|---|---|---|---|---|
| public_account | -0.0728 | 0.0189 | -3.8568 | < 0.0001 |
| private_account | -0.3897 | 0.0218 | -17.8795 | < 0.0001 |
| round_donations_others | 0.1503 | 0.0568 | 2.6442 | 0.0082 |
| actions_prev | 0.0886 | 0.1039 | 0.8532 | 0.3936 |
| GameHalf2ndHalf | -12.0490 | 0.5227 | -23.0520 | < 0.0001 |
| rnd | -0.1145 | 0.1614 | -0.7098 | 0.4779 |
| public_account:GameHalf2ndHalf | 0.0347 | 0.0192 | 1.8070 | 0.0708 |
| round_donations_others:GameHalf2ndHalf | 0.0121 | 0.0621 | 0.1956 | 0.8449 |
| actions_prev:GameHalf2ndHalf | 0.2040 | 0.1210 | 1.6852 | 0.0920 |
| private_account:GameHalf2ndHalf | 0.3167 | 0.0224 | 14.1248 | < 0.0001 |

| Intercepts | | | | |
|---|---|---|---|---|
| 0\|2 | -15.4816 | 0.5294 | -29.2444 | < 0.0001 |
| 2\|4 | -13.6380 | 0.5254 | -25.9587 | < 0.0001 |

| | |
|---|---|
| Residual Deviance: | 1145.2530 |
| AIC: | 1169.2530 |

Table S4. Ordered linear regression for the treatment with high uncertainty. Similar to the low uncertainty case, for high uncertainty we can observe that both time and the donations of other in the group are relevant features.

**Polynomial fitting**

Our analysis until now was mostly focused on identifying linear correlations, and their sign, between participants' contributions and the contributions of their group mates. This choice allows us to extract meaningful relationships while avoiding overfitting. However, in some of the studied cases, as we would expect, the data dependency might be represented better by a curvilinear/polynomial model. Below we show that the conclusions presented in the main manuscript remain valid if high order fitting is chosen.

In Tables S5-S10, we display the results of an ANOVA test that compares polynomial regressions of different order (from 1 to 4) for all the 6 cases analyzed in Figure 3 of the main manuscript. This test indicates whether increasing the order of the polynomial regression issues a significant improvement. In only 3 cases, a polynomial fit is significantly better than the linear model: no uncertainty and players meeting the target; low uncertainty and not meeting the target; and high uncertainty and not meeting the target. In Figure S6, we show results analogous to Figure 3 of the main manuscript, but using the model that fits the best each case. We show that our conclusions do not change, and perhaps, it is even clearer that, in the certainty case, when players meet the target, they adopt a slightly compensatory behavior, while considerably lowering their contributions when the rest of the group adopt extreme actions: they donate either too much or too little.

| Model | Res.Df | RSS | Df | Sum of Sq. | F | Pr(>F) |
|---|---|---|---|---|---|---|
| 1 | 6 | 3.6549 | | | | |
| 2 | 5 | 2.3961 | 1 | 1.25873 | 3.1757 | 0.1728 |
| 3 | 4 | 1.867 | 1 | 0.52912 | 1.3349 | 0.3316 |
| 4 | 3 | 1.1891 | 1 | 0.67793 | 1.7104 | 0.2821 |

Table S5. ANOVA test for the no uncertainty treatment, and players that did not meet the target. No higher order polynomial model performs significantly better than the linear fit.

| Model | Res.Df | RSS | Df | Sum of Sq. | F | Pr(>F) | |
|---|---|---|---|---|---|---|---|
| 1 | 8 | 3.3276 | | | | | |
| 2 | 7 | 3.2601 | 1 | 0.06752 | 0.6832 | 0.44611 | |
| 3 | 6 | 3.2579 | 1 | 0.00219 | 0.0222 | 0.887473 | |
| 4 | 5 | 0.4941 | 1 | 2.76378 | 27.9654 | 0.003223 | ** |

Table S6. ANOVA test for the no uncertainty treatment, and players that met the target. A fourth order polynomial generates a significantly better fit than the linear model.

| Model | Res.Df | RSS | Df | Sum of Sq. | F | Pr(>F) | |
|---|---|---|---|---|---|---|---|
| 1 | 6 | 2.37502 | | | | | |
| 2 | 5 | 2.277 | 1 | 0.09802 | 0.377 | 0.58265 | |
| 3 | 4 | 0.78189 | 1 | 1.49511 | 5.75 | 0.09605 | . |
| 4 | 3 | 0.78006 | 1 | 0.00183 | 0.007 | 0.9385 | |

Table S7. ANOVA test for the low uncertainty treatment, and players that did not meet the target.  A third order polynomial provides a better fit than the linear model.

| Model | Res.Df | RSS | Df | Sum of Sq. | F | Pr(>F) |
|---|---|---|---|---|---|---|
| 1 | 9 | 6.7944 | | | | |
| 2 | 8 | 6.6108 | 1 | 0.18354 | 0.2259 | 0.6514 |
| 3 | 7 | 5.6696 | 1 | 0.94127 | 1.1584 | 0.3232 |
| 4 | 6 | 4.8753 | 1 | 0.79422 | 0.9774 | 0.361 |

Table S8. ANOVA test for the low uncertainty treatment, and players that met the target. No model provides a better fit than the lineal model.

| Model | Res.Df | RSS | Df | Sum of Sq. | F | Pr(>F) | |
|---|---|---|---|---|---|---|---|
| 1 | 7 | 2.22619 | | | | | |
| 2 | 6 | 1.49467 | 1 | 0.73152 | 5.9811 | 0.07078 | . |
| 3 | 5 | 0.93891 | 1 | 0.55576 | 4.5441 | 0.10002 | |
| 4 | 4 | 0.48922 | 1 | 0.44969 | 3.6768 | 0.12764 | |

Table S9. ANOVA test for the high uncertainty treatment, and players that did not meet the target. A second order polynomial provides a better fit than the linear model.

| Model | Res.Df | RSS | Df | Sum of Sq. | F | Pr(>F) |
|---|---|---|---|---|---|---|
| 1 | 7 | 3.6 | | | | |
| 2 | 6 | 3.5877 | 1 | 0.01231 | 0.0191 | 0.8969 |
| 3 | 5 | 3.2477 | 1 | 0.33999 | 0.5262 | 0.5084 |
| 4 | 4 | 2.5846 | 1 | 0.66316 | 1.0263 | 0.3683 |

Table S10. ANOVA test for the high uncertainty treatment, and players that met the target. No model provides a better fit than the linear model.

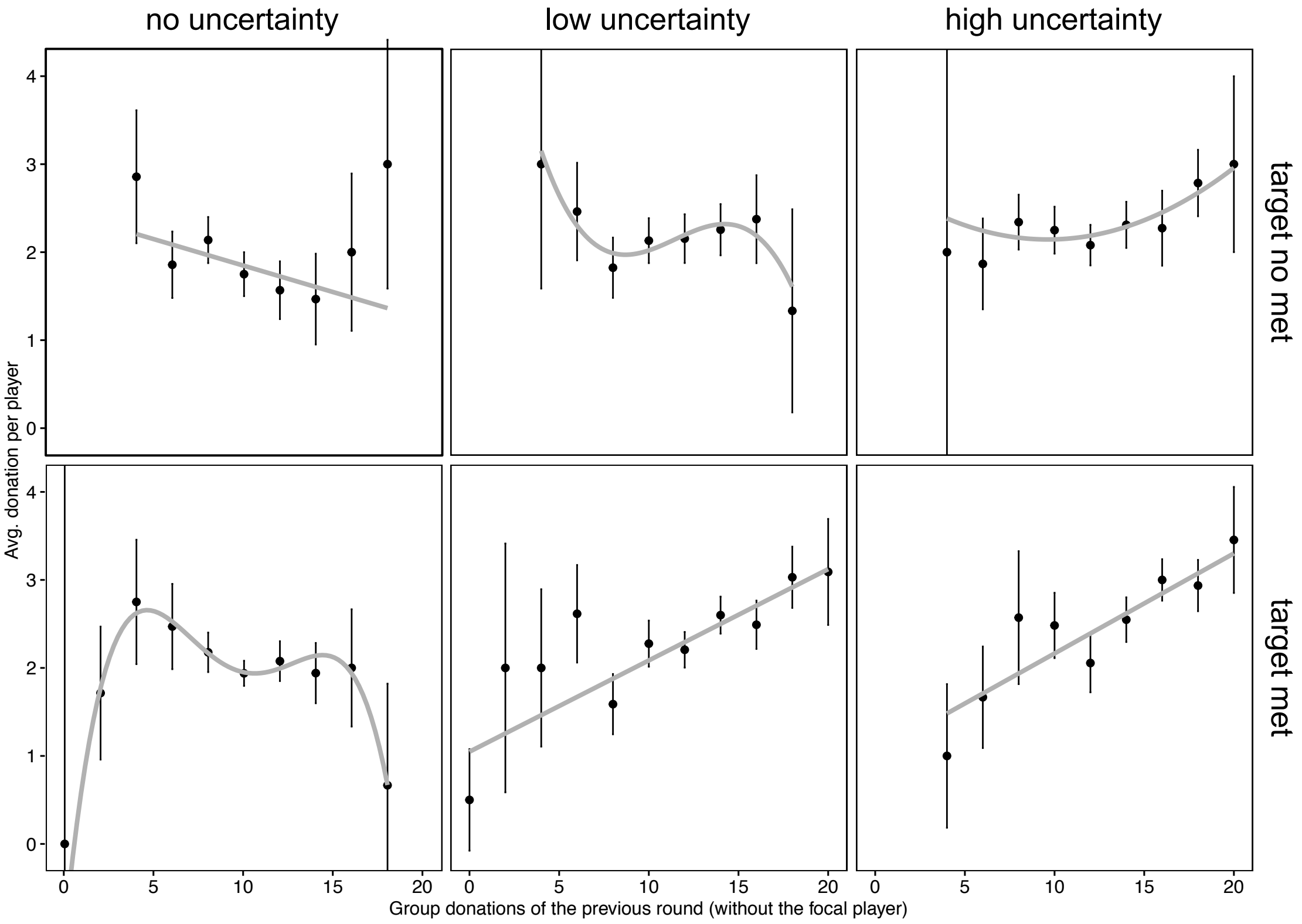

**Figure S6. Analysis of conditional behavior using the best fitting model for each case.** Applying the best fitting model identified through our tests to each treatment does not change our conclusions, but highlights that on the certainty treatment, when players meet the collective target, they display a slight compensatory behavior, while considerably lowering their contributions when the rest of the group adopt extreme actions: they donate either too much or too little.

## Supplementary model results

Here we show additional results of our evolutionary game theoretical model[2–4]. As fully detailed in the Methods section, we consider five representative strategies encountered in our experiments: *always-0 (**AL0**)*, *always-2 (**AL2**)*, *always-4 (**AL4**)*, *compensator (**COMP**)*, and *reciprocal (**RECI**)*. The first three strategies always invest the same: 0, 2, and 4, respectively, independently of the behavior of their opponents. *Compensators* only invest when the group members did not invest, and *reciprocal* players will invest as long as the group members are always investing (see Methods for details). All strategies stop investing once the threshold is reached.

We analyze the behavioral dynamics in large (but finite) populations, when individuals revise their choice through imitation dynamics or social learning[5–7].

Strategies that do well are imitated more often and spread in the population. Particularly, at each time step, the strategy of one randomly selected individual $A$ is updated. With probability μ, $A$ undergoes a mutation. With probability $1 - \mu$, another randomly selected individual $B$ acts as a role model for $A$. The probability that $A$ adopts the strategy of $B$ equals $\mathrm{p} = \left[1 + e^{\beta(f_A - f_B)}\right]^{-1}$; $A$ sticks to his/her former strategy with probability $1 - \mathrm{p}$. This update rule is known as the pairwise comparison rule [7] where $f_A$ and $f_B$ denote the fitness of individual $A$ and $B$, respectively, as defined in the Methods section. The parameter $\beta \geq 0$, measures the strength of selection. In the limit of strong selection ($\beta \rightarrow \infty$), the probability $p$ is either zero or one. In the limit of weak selection ($\beta \rightarrow 0$), $p$ is always equal to $1/2$, irrespective of the fitness of $A$ and $B$. We have checked that our results are qualitatively invariant for a broad range of values of β.

If the mutation probability μ is sufficiently small, the time between two mutation events is so large that, before the next mutation occurs, the population will evolve to a configuration in which all individuals adopt the same strategy. In this case, the dynamics can be approximated by means of a Markov chain whose states correspond to the five homogeneous states of the population (for each of the five strategies studied here).

The transitions of this Markov chain are defined by the fixation probabilities $\rho_{ij}$ of a single mutant with strategy j in a population of individuals adopting another strategy i. The transition matrix $\Lambda = \left[\Lambda_{ij}\right]$ combines the different probabilities that a population in a homogeneous state $S_i$ will end up in state $S_j$ after the occurrence of one single mutation. This matrix is given by $\Lambda_{ij} = \frac{\rho_{ij}}{4} (j \neq i)$, whereas the diagonal of the transition matrix is defined by $\Lambda_{ii} = 1 - \frac{1}{4}\sum_{j \neq i} \rho_{ij}$ (note that 4 is the number of strategies minus one). The normalized left eigenvector associated with eigenvalue 1 of matrix Λ determines the stationary distribution, i.e, the fraction of time the population spends in each of the homogeneous states.

In the limit of neutral selection ($\beta = 0$), the fixation probabilities become independent of the fitness values and equal to $1/\mathrm{Z}$, offering a convenient reference scenario.

Transitions above this threshold are said to be favored by natural selection. In Figure S6 we show these transitions for two of the most paradigmatic scenarios: NU and HU. For instance, if an arrow goes from a state with strategy $i$ to state with strategy $j$, it indicates that a mutant of strategy $j$ will invade the population of strategy $i$ with a probability which is higher than the one we obtain from neutral drift. The absence of an arrow indicates that such transition will occur with a low probability, i.e., with a probability lower than $1/Z$. In this context, a strategy $j$ is said to be evolutionary robust (ERS)[3,4] if no mutant, adopting any other strategy, has a selective advantage. In other words, we can identify strategies that are evolutionary robust (a measure of stability) by noticing that there is no arrow emerging from its respective node (see Figure S6).

Figure S6 illustrates a reference scenario in what concerns the invasion dynamics of strategies. In the absence of uncertainty, *always-0* and the *always-2* strategy are the only two evolutionary robust strategies. Moreover, given the number and strength of the transitions, the fair strategy can easily become the most prevalent behavior. Differently, under high timing uncertainty, the *always-2* strategy starts to be invaded by *compensators* and *always-0* (Figure S7b), changing the ecology of behaviors observed in the absence of uncertainty. In fact, uncertainty can easily lead to complex behavioral dynamics, with cyclic dominances, and no evolutionary robust strategies, as illustrated in the right panel of Figure S6. *Reciprocators* can invade *always-0*, yet losing to compensators, which, in turn, are invaded by *always-0*. From this cyclic dynamic, even if not stable, both conditional strategies emerge as prevailing strategies, leading both to the emergence of reciprocity and polarization (see main text).

The limit of rare mutations allows us to conveniently employ a small-scale Markov chain to analytically compute the prevalence of each strategy. This is achieved by restricting the number of strategies simultaneously co-existing in a population (and groups) to a maximum of two. However, for arbitrary mutation rates, we may have a complex co-existence of more than two strategies, calling for the adoption of large-scale computer simulations to confirm the validity of our theoretical results in other mutation (or exploration) regimes. To perform these computer simulations, we mimic the evolutionary process described above, with discrete steps involving imitation and mutation, yet without any constraint in the value of $\mu$ (here a free variable). At the beginning of each simulation, each individual randomly adopts one of the five strategies. In each generation, $Z$ individuals are chosen to revise their strategy (in an asynchronous manner). For each combination of parameters, we run 30 simulations, each lasting $10^9$ generations. The fitness of each individual A is calculated as the average return earned from $10^3$ games played against N-1 individuals randomly selected from the population. The group achievement is computed from the average fraction of groups that surpassed the threshold after a transient period of $10^5$ generations. The same criterion is used to compute the overall level of polarization emerging from each simulation.

In Figure S8, we confirm that the stationary distribution obtained under the small mutation limit assumption is valid for a wide range of mutation values. Additionally, we also include the polarization results, considering only group combinations that achieve the target, with and without mutation (see Figure S9), showing that results discussed in the main text remain valid for a broad interval of exploration rates.

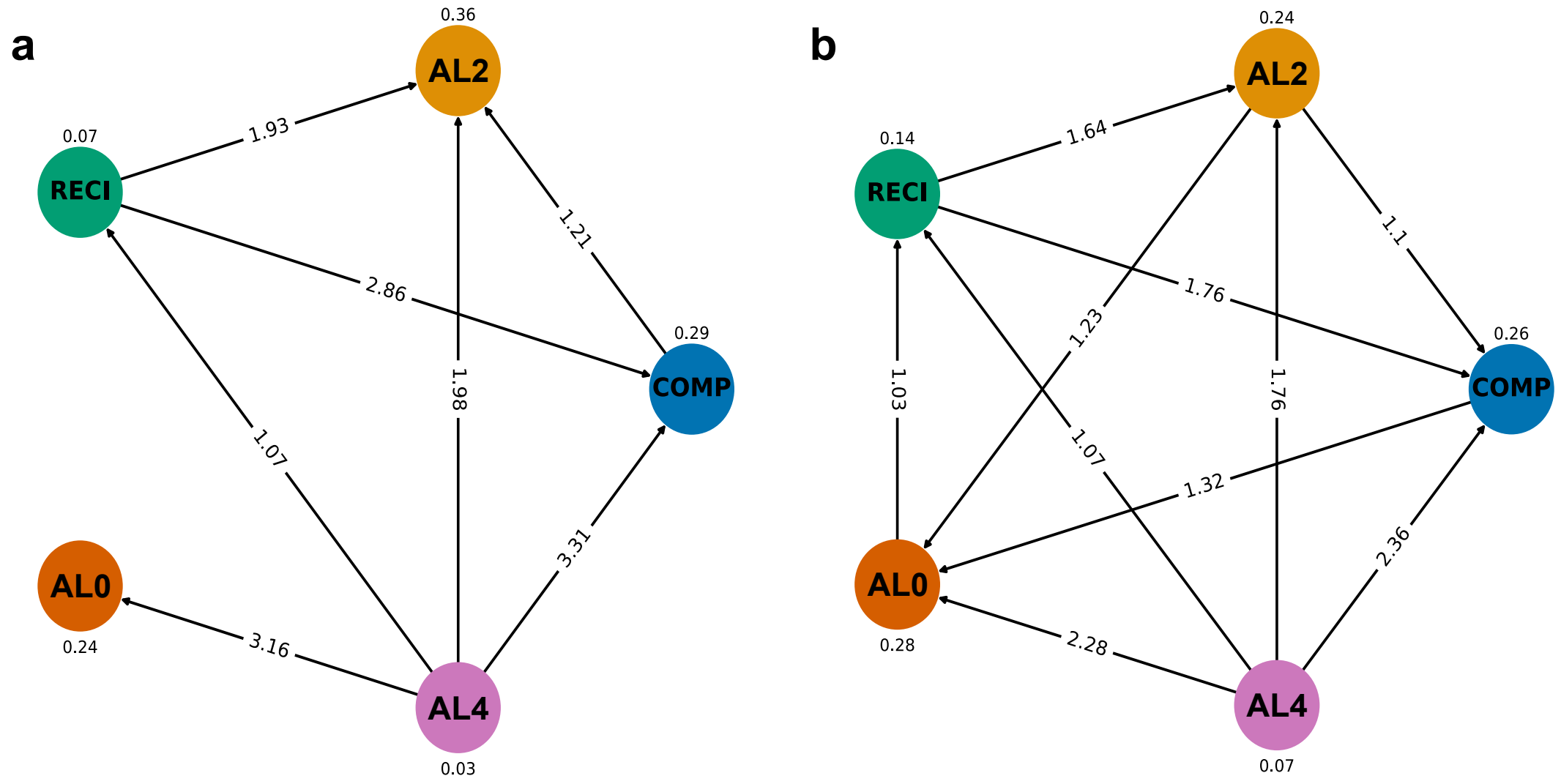

Figure S7. Markov chain depicting the transition probabilities between states. In panel a) we show the Markov chain when there is no timing uncertainty. An arrow that goes from state $\boldsymbol{i}$ to $\boldsymbol{j}$ indicates that a population in state $\boldsymbol{i}$ (where all members of the population are of strategy $\boldsymbol{i}$) will transition to strategy $\boldsymbol{j}$ with a probability higher than random drift. The number on top of each state indicates the stationary distribution,

i.e., the time the population spends in that state. When there is no timing uncertainty, *always-2* is an evolutionary stable strategy (all arrows point to it, and none goes out). However, in the high timing uncertainty case (Panel b), it becomes dominated by *always-0*. Also, *reciprocal* strategy weakly dominates *always-0*, resulting in a cyclic dynamic (no strategy dominates). This explain the reduction in *always-2* players and the increase of *reciprocals*.

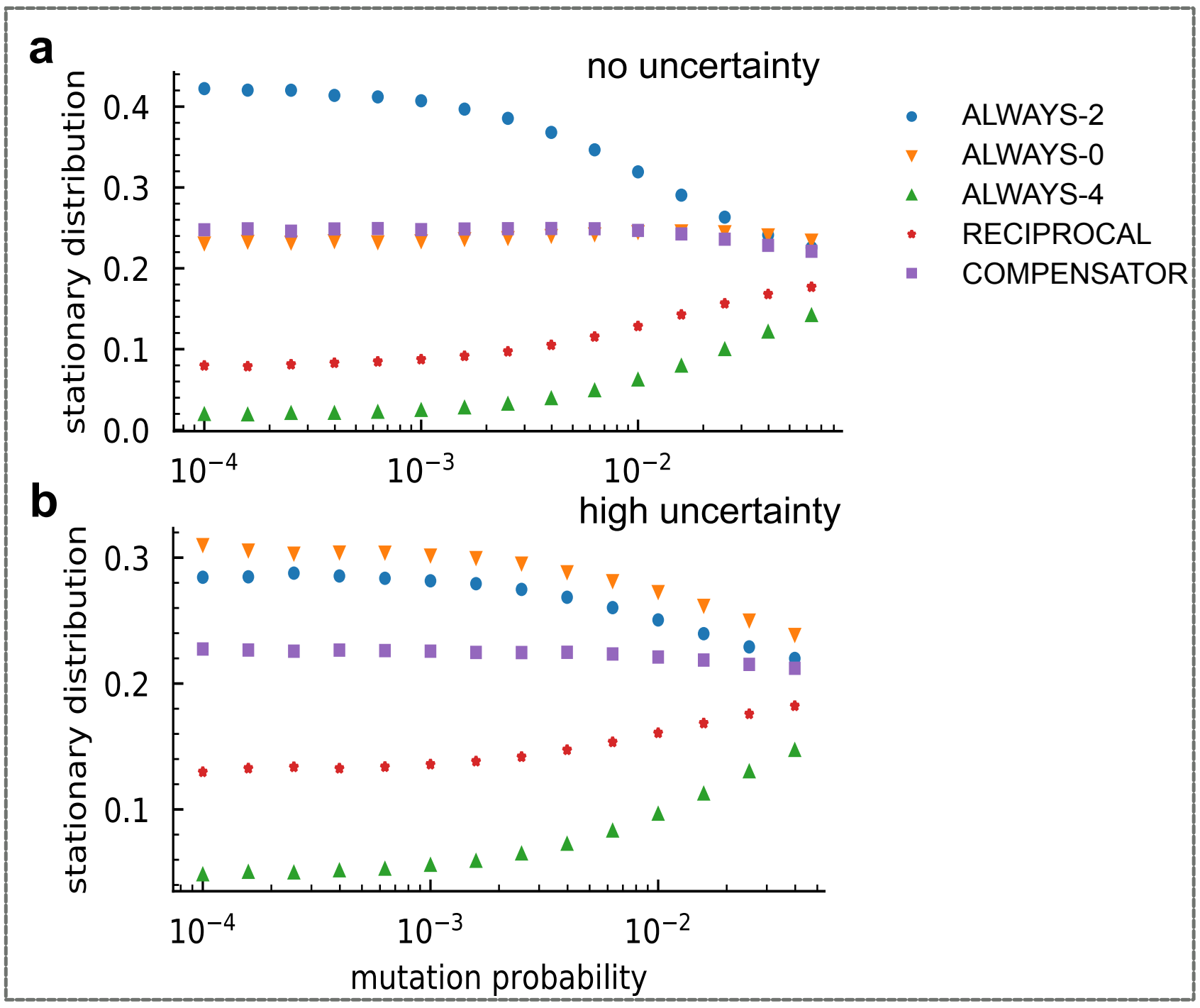

Figure S8. Influence of noise (mutation) in the stationary distribution. This figure shows how the stationary distribution of the monomorphic states (only one strategy in the population) is affected by the mutation probability (the probability that a random mutant appears in the population). The results show that the small mutation assumption in the theoretical model is valid for a large value of mutation probabilities both in the case of no uncertainty (Panel a) and high uncertainty (Panel b). Only for mutation values higher than $\mathbf{10^{-2}}$ does the distribution get significantly affected. In this case, mixed states (more than one strategy survives in the population) become more common ($\boldsymbol{\beta = 0.003}, \boldsymbol{Z = 50}$).

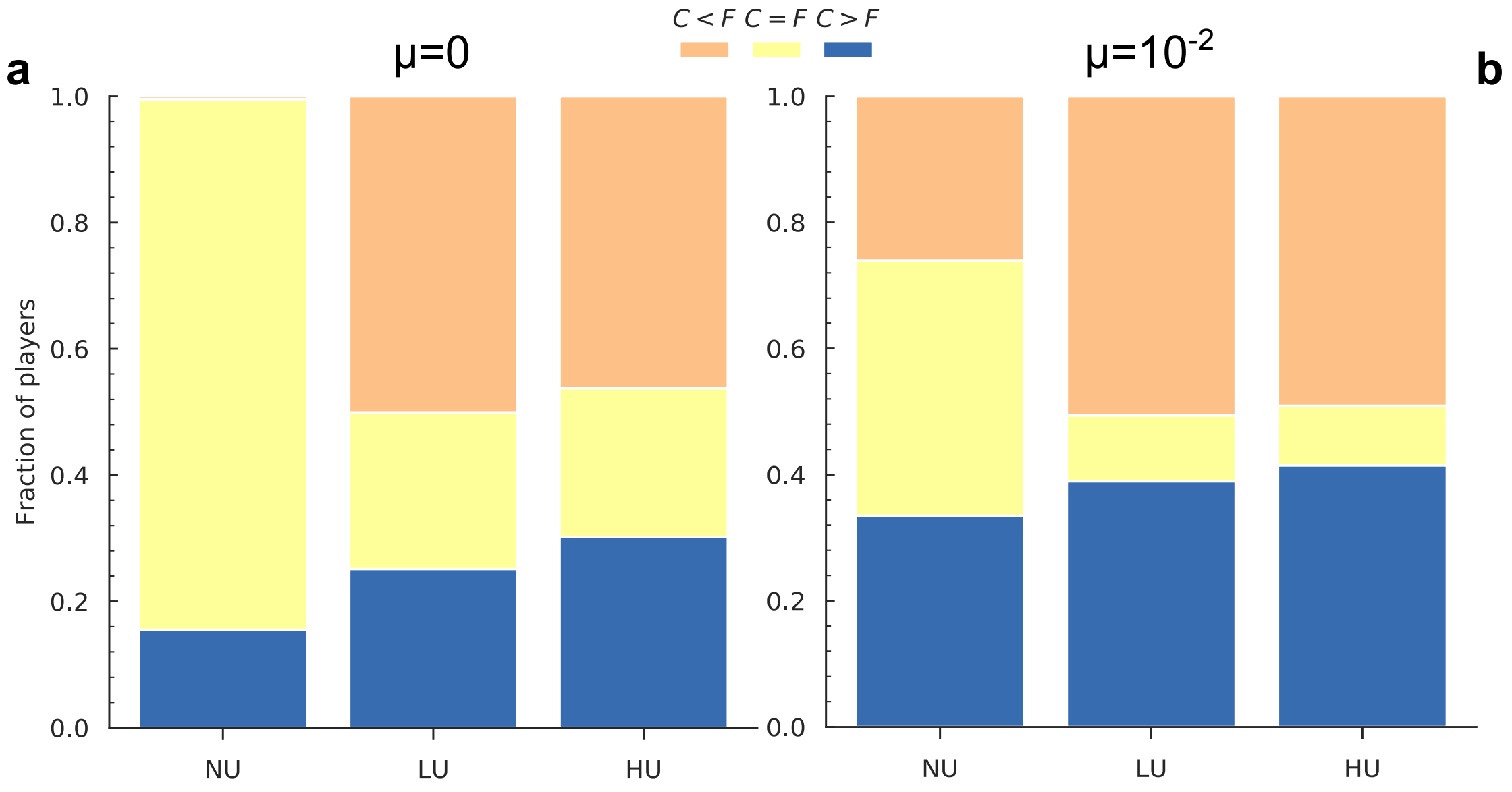

Figure S9. Distribution of donations when only considering successful groups. In panel (a) we show the fraction of the population that invests more, less or equal to the fair donation (*F*), considering only groups that achieve the target. We can observe that, while in the no uncertainty (NU) case, successful groups either invest equal or above F, an important fraction of the population invests below F in the timing uncertainty cases. This indicates the emergence of polarization with timing uncertainty, as we observed in the experiments. In panel (b), we show the same results, but considering a mutation rate $\mu = 10^{-2}$. Here, our model' results matches closely the experimental results for NU, while it predicts an even more extreme case of polarization for LU and HU.